\documentclass[12pt]{iopart}

\usepackage{iopams}  
\usepackage{graphicx}
\usepackage{color}
\usepackage{ulem}
\newcommand{\del}[1]{}

\begin{document}

\title[Morse index and bifurcation for figure-eight choreographies]
{Morse index and bifurcation for figure-eight choreographies of the equal mass three-body problem}

\author{Hiroshi Fukuda$^1$, Toshiaki Fujiwara$^1$ and Hiroshi Ozaki$^2$}

\address{$^1$ College of Liberal Arts and Sciences, Kitasato University, 1-15-1 Kitasato, Sagamihara, Kanagawa 252-0329, Japan}
\address{$^2$ Laboratory of general education for science and technology, Faculty of Science, Tokai University, 
4-1-1 Kita-Kaname, Hiratsuka, Kanagawa, 259-1292, Japan}
\ead{fukuda@kitasato-u.ac.jp, fujiwara@kitasato-u.ac.jp and ozaki@tokai-u.jp}
\vspace{10 pt}
\begin{indented}
\item[] \today %8 August 2017
\end{indented}

\begin{abstract}
We report on  the Morse index and periodic solutions bifurcating from 
the figure-eight choreography for the equal mass three-body problem 
under homogeneous potential $-1/r^a$  for $a \ge 0$,  
and under Lennard-Jones (LJ) type potential $1/r^{12}-1/r^6$,
where $r$ is a distance between bodies.
It is shown that the Morse index changes at a bifurcation point and 
all solutions bifurcating are approximated by variational functions 
responsible for the change of the Morse index.
Inversely we observed %numerically 
bifurcation occurs at every point where the Morse index changes 
for the figure-eight choreography under $-1/r^a$, 
and for $\alpha$ solution under LJ type potential, 
where $\alpha$ solution is a figure-eight choreography 
tending to that under $-1/r^6$ for infinitely large period. 
Thus, to our numerical studies, 
change of the Morse index is not only necessary but also sufficient condition for bifurcation
for these choreographies. 
Further we observed that 
the change of the Morse index is equal to 
the number of bifurcated solutions regarding solutions with congruent orbits as the same solution.
\end{abstract}

%\noindent{\it Keywords}: X, Y, Z
\submitto{{\it J. Phys. A: Math. Theor.}}
%\maketitle
% 
% For two-column output uncomment the next line and choose [10 pt] rather than [12 pt] in the \documentclass declaration

\section{Introduction}
\label{sec:intro}
Choreographic motion of $n$ bodies is a periodic motion on a closed orbit, 
$n$ identical bodies chase each other on the orbit with equal time-spacing. 
Moore \cite{moore} found a remarkable figure-eight choreographic solution for $n=3$ 
under homogeneous potential $-1/r^a$ by numerical calculations, 
where $r$ is a distance between bodies. 
Chenciner and Montgomery \cite{chenAndMont} gave a mathematical proof 
of its existence for $a=1$
by variational method. 
The detailed initial conditions for three bodies are found in \cite{chenAndMont,simoH}.

%After that, 
Sbano \cite{sbano2005}, and Sbano and Southall \cite{sbano},
studied $n$-body choreographic solutions 
under an inhomogeneous potential %in the form
\begin{equation}
   u^{LJ}(r)=\frac{1}{r^{12}}-\frac{1}{r^6},
\label{eq:LJu}
\end{equation}
a model potential between atoms called Lennard-Jones-type (hereafter LJ) potential.
Sbano and Southall \cite{sbano} proved that 
there exist at least two $n$-body choreographic solutions for sufficiently large period, 
and there exists no solution for small period. 
We confirmed their theorem numerically for $n=3$ and 
unexpectedly found a multitude of three-body figure-eight choreographic solutions
under LJ-type potential (\ref{eq:LJu})  \cite{fukuda}.

%Then, 
Following Shibayama's preliminary calculation for $a=1$ \cite{shibayama},
we did accurate numerical calculation of the Morse index,  
for the three-body figure-eight choreography
in the domain of periodic function \cite{fukuda2}. 
Here the Morse index is a number of independent variational functions 
giving negative second variation of action functional. 

In our paper \cite{fukuda2}, 
a strong relationship between the Morse index
%, bifurcation of figure-eight choreography 
and H solution found by Sim\'{o} \cite{simoH},
which is a periodic solution close to the figure-eight choreography but made up of three distinct orbits, 
was suggested. 
On the other hand Gal\'{a}n \etal \cite{Vanderbauwhede}
showed that the H solution bifurcated from figure-eight choreography 
by changing the masses of three bodies. 
They also found many different periodic orbits on figure-eight  \cite{Munoz-Almaraz}.

There are several researches on the Morse index for periodic solution of three-body problem.
Barutello \etal \cite{barutello} calculated the Morse index mathematically 
for the Lagrangian circular solution, and 
Hu and Sun \cite{hu, hu2} for elliptic Lagrangian solutions, to discuss the linear stability.

In this paper, we show a relationship between 
the Morse index of the figure-eight choreographies and 
periodic solutions bifurcating %from the figure-eight choreographies 
for a system of three identical bodies interacting through a homogeneous potential 
or through LJ-type potential (\ref{eq:LJu}). 
In section \ref{sec:morse}, 
we  show the Morse index changes at a bifurcation point
and solutions bifurcating are approximated by variational functions 
responsible for change of the Morse index.

In section \ref{sec:homo}, 
we discuss the bifurcation from the figure-eight choreography  
under homogeneous potential, $-1/r^a$, by changing $a$. 
In section \ref{sec:simoH}, 
we show the H solution bifurcates 
% under homogeneous potential 
at $a=0.9966$ where the Morse index changes.
In section \ref{sec:D}, 
another bifurcation at $a=1.3424$  is shown.
These bifurcations at $a=0.9966$ and $1.3424$ are first found in 2005
by Mu\~{n}oz-Almaraz \etal \cite{Vanderbauwhede_mail,Munoz-Almaraz2} 
using AUTO \cite{AUTO1,AUTO2}. 
There is no other point where the Morse index changes for $a \ge 0$. 

In section~\ref{sec:LJ}, 
we discuss the bifurcation of the $\alpha$ solution 
for the system under LJ-type potential, 
%which tends to the figure-eight choreography under $-1/r^6$ potential
%for infinitely large period.
where the $\alpha$ solution is a figure-eight choreography 
tending to that under $-1/r^6$ for infinitely large period. 
There are seven points where the Morse index changes for the $\alpha$ solution. 
In section~\ref{sec:LJH}
we show four points bifurcate periodic but non choreographic solutions, 
which are 
the same type of the bifurcations discussed in section~\ref{sec:homo}. 
In section~\ref{sec:LJC},
we show the rest three points 
yield choreographic solutions less symmetric than figure-eight. 
Alain Chenciner, in ICM 2002, asked a question about the existence of 
less symmetric figure-eight \cite{chenciner,shibayama_mail}. 
We can say ``Yes", they exist under LJ-type potential. 
Section~\ref{sec:summary} is a summary and discussions.

Our numerical results in this paper were calculated by Mathematica~11.1.

\section{Morse index and bifurcation}
\label{sec:morse}

\subsection{Morse index and eigenvalue problem}
\label{sec:eigen}
For a system of three identical bodies in classical mechanics, 
we consider periodic solutions to equations of motion, 
\begin{equation}\label{eq:motion}
	\frac{d}{d t} \frac{\partial L}{\partial \dot{q}_i} = \frac{\partial L}{\partial q_i},
	\; i=1,2, \ldots, 6,
\end{equation}
where dot represents a differentiation in $t$.
$L$ is the Lagrangian with the potential energy $U(q)$, 
\begin{equation}\label{eq:L}
	L%(q,\dot{q})
	=\sum_{i=1}^{6}\frac{\dot{q}_i^2}{2}-U(q),   
\end{equation}
and 
\begin{equation}\label{key}
	q(t)=(q_{1}(t), q_{2}(t),\ldots, q_{6}(t))^*
\end{equation}
a six component vector composed of position vectors 
\begin{equation}\label{key}
	\bi{r}_b(t) = (x_{b}(t), y_{b}(t))^* = (q_{2 b-1}(t), q_{2 b}(t))^*
\end{equation}
for body $b=1,2,3$ 
%
%moving 
in a plane, 
%dot represents a differentiation in $t$,
where $^*$ represents transpose.
The subscript of six component vector is assumed to be in the range
between 1 and 6. %with translation by 6.
% and 
%also the subscript for body is assumed to be in the range 1 and 3 
%with translation by 3.

For a periodic solution $q(t+T)=q(t)$ 
and variation function $\delta q(t+T)=\delta q(t)$ 
%is a variation function 
%$\delta q(t+T)=\delta q(t)$,
with period $T$,
%we calculate 
the Morse index $N$ is defined as a number of independent  variation function $\delta q(t)$ 
which make the second variation $S^{(2)}$ with
\begin{equation}\label{S^(k)}
	S^{(k)}=\int_{0}^{T}d t \left( \sum_{i}(\delta q_i \frac{\partial}{\partial q_i}+
	\dot{\delta q}_i \frac{\partial}{\partial \dot{q}_i}) \right) ^k L
\end{equation}
%with $k=2$ 
of the action functional
\begin{equation}\label{key}
	S(q)=\int_{0}^{T} L(q,\dot{q}) d t
\end{equation}
negative.
%The $k$'th variation $S^{(k)}$ of the action $S(q)$ is defined as the $k$'th coefficients in
%\begin{equation}\label{eq:S}
%S(q+h \delta q)=S^{(0)}+h S^{(1)}+\frac{h^2}{2!} S^{(2)}+\cdots,
%\end{equation}
%thus
%\begin{equation}\label{key}
%S^{(k)}=\int_{0}^{T}d t \left( \sum_{i}(\delta q_i \frac{\partial}{\partial q_i}+
%\dot{\delta q}_i \frac{\partial}{\partial \dot{q}_i}) \right) ^k L,
%\end{equation}
%where 
%$h$ is a real number and 
%$\delta q$ is a variation function with period $T$,
%$\delta q(t+T)=\delta q(t)$,
%
%By partial integration, the second variation 
Since the $S^{(2)}$ is written as
\begin{equation}\label{key}
%S^{(2)}=\int_{0}^{T} d t {\delta q}^* \hat{H} \delta q
S^{(2)}=(\delta q, \hat{H} \delta q)
\end{equation}
by $6\times 6$ matrix operator $\hat{H}$, 
\begin{equation}\label{def:H}
	(\hat{H})_{i j} = -\delta_{i j} \frac{d^2}{d t^2} -\frac{\partial^2 U}{\partial q_i \partial q_j}, 
\end{equation}
%with 
%\begin{equation}\label{eq:Umat}
%U_{i j}(t) 
%%= \frac{\partial^2 L}{\partial q_i \partial q_j} 
%= \frac{\partial^2 U}{\partial q_i \partial q_j},
%\end{equation}
the Morse index is a number of negative eigenvalues of 
the eigenvalue problem, 
\begin{equation}\label{eq:eigen}
	\hat{H} \psi = \lambda \psi, 
\end{equation}
\begin{equation}\label{psi:T}
	\psi(t+T)=\psi(t),  
\end{equation}
\begin{equation}\label{eq:norm}
	(\psi,\psi)=1,  
\end{equation}
with the second variation 
\begin{equation}\label{key}
S^{(2)}=\lambda
\end{equation}
and the variation function 
\begin{equation}\label{dq=psi}
	\delta q=\psi, 
\end{equation}
where $(f,g)$ is the inner product defined by 
\begin{equation}\label{key}
	(f,g)=\int_{0}^{T} d t f^* g
\end{equation}
and $\delta_{i j}$ the Kronecker delta. 
%and the eigenfunction $\psi$ is assumed to be normalized as 

\subsection{Bifurcation and eigenvalue problem}
\label{sec:bifurcation}

For a periodic solution of equation of motion (\ref{eq:motion})
with some parameter $\xi$, $q(t; \xi)$,
suppose the other periodic solution $q^b(t; \xi)$ 
bifurcates at $\xi=\xi_0$, that is,
\begin{equation}\label{key}
	\lim_{\xi \to \xi_0} q^b(t; \xi) = q(t; \xi_0). 
\end{equation}
Since for the Lagrangian (\ref{eq:L})
\begin{equation}\label{eq:q}
	\ddot{q}=-\frac{\partial U}{\partial q_i}
\end{equation}
and 
\begin{equation}\label{eq:qb}
	\ddot{q}^b=-\frac{\partial U}{\partial q^b_i}, 
\end{equation}
$\Delta q=q^b-q$ satisfies 
\begin{equation}\label{key}
	\frac{d^2}{d t^2} \Delta q_i 
	= -\left.\frac{\partial U}{\partial q_i}\right|_{q=q^b} + \frac{\partial U}{\partial q_i}
	= -\sum_j{\frac{\partial^2 U}{\partial q_j \partial q_i}\Delta q_j} + O(|\Delta q|^2),
\end{equation}
that is, 
\begin{equation}\label{Hdq=0}
	\hat{H} \Delta q  \to 0 \mbox{ for } \xi \to \xi_0. 
\end{equation}
Thus equation (\ref{Hdq=0}) is a necessary condition for $\xi_0$ to be a bifurcation point.
%yielding the solutions with the same period $T$

%This means 
When $\xi \to \xi_0$, equation (\ref{Hdq=0}) shows that  
%since $\Delta q \to 0$,  
%the bifurcating function $q^b$ is written by 
$\Delta q$ goes to the eigenfunction of $\hat{H}$ %defined in (\ref{def:H}) 
whose eigenvalue $\lambda$ is zero.
%
%\begin{equation}\label{eq:q+psi}
%	q^b \to q+\psi.
%\end{equation} 
%
In other words, at a bifurcation point 
some eigenvalue $\lambda$ of $\hat{H}$ has to go to zero.
%$\lambda \to 0$.
%Since eigenvalue problem is defined in the periodic function with period $T$,
%as a corollary of (\ref{Hdq=0}),
%a necessary condition for $\xi_0$ to be a bifurcation point 
%yielding the solutions with the same period $T$
%is that some eigenvalue $\lambda$ of $\hat{H}$ has to go to zero,
%$\lambda \to 0$ at $\xi_0$.
%
Thus, using the normalized eigenfunctions 
$\psi^{(k)}$
%, $k=1, 2, \ldots, g$, 
of the eigenvalue $\lambda \to 0$ 
%near the bifurcation point,
we have an approximate expression by variated orbit 
\begin{equation}\label{q+psi}
	Q = q + h \sum_k^{g} c_k \psi^{(k)}
\end{equation}
for the bifurcating solution
\begin{equation}\label{q^b->q+psi}
	q^b = q+\Delta q \to Q \mbox{ for } \xi \to \xi_0, 
\end{equation}
where $g$ is a degeneracy of the $\lambda$, 
$c_k$ and $h$ are real coefficients with $\sum_k^g c_k^2 = 1$ and $h \to 0$.
Since the variated orbit $Q$ has its own symmetry independent of $\xi$ and $h$ \cite{fukuda2},
bifurcating solution $q^b$ will be found within the symmetry.

\subsection{Morse index and bifurcation}
\label{sec:deltaN}

We define change of the Morse index $N(\xi)$ by 
\begin{equation}\label{DeltaN:def}
	\Delta N(\xi) = \lim_{\xi' \to \xi+0} N(\xi') - \lim_{\xi' \to \xi-0} N(\xi'). 
\end{equation}
Thus 
\begin{equation}\label{key}
	g=|\Delta N(\xi_0)|
\end{equation}
and an eigenvalue $\lambda(\xi)$ changes the sign at $\xi=\xi_0$,  
%\begin{equation}\label{lambda:derivative}
%	\Delta N(\xi_0)  \frac{d\lambda}{d\xi} \le 0,  
%\end{equation}
%or
%
%Since $\lambda(\xi_0)=0$, (\ref{lambda:derivative}) is written as 
\begin{equation}\label{lambda:behaviour}
	\Delta N(\xi_0)(\xi-\xi_0) \lambda(\xi) \le 0 
\end{equation}
around $\xi_0$. 

Then the sign of the action, 
\begin{equation}\label{DeltaS:def}
	\Delta S(\xi)=S(q^b(t;\xi))-S(q(t;\xi)), 
\end{equation}
has the same sign as $\lambda(\xi)$ 
\begin{equation}\label{lambdaDeltaS}
	\Delta S(\xi) \lambda(\xi)  \ge 0
\end{equation}
and 
%is also given by $\Delta N (\xi_0)$ as (\ref{lambda:behaviour}) 
\begin{equation}\label{DeltaS:behaveror}
	\Delta N(\xi_0) (\xi-\xi_0) \Delta S(\xi) \le 0 
\end{equation}
around $\xi_0$.  
Since 
$S$ is expanded in $h$ with coefficients (\ref{S^(k)}) as 
\begin{equation}\label{eq:S}
	S(q+h \delta q)=S^{(0)}+h S^{(1)}+\frac{h^2}{2!} S^{(2)}+\cdots,
\end{equation}
and $q^b$ tends to $Q=q+h \delta q$ with $\delta q=\sum_k^g c_k \psi^{(k)}$ 
for $\xi \to \xi_0$, 
% in (\ref{S^(k)}), 
\begin{equation}\label{DeltaS:derivation}
	\Delta S(\xi) 
	\to S(Q)-S(q) 
	= \frac{h^2}{2} S^{(2)} + \frac{h^3}{3!} S^{(3)} + \cdots 
	= \frac{h^2}{2} \lambda  + O(h^3).  
\end{equation}
%where $h \to 0$.
%$q+h\sum_k^g c_k \psi^{(k)}$
Thus (\ref{lambdaDeltaS}) then (\ref{DeltaS:behaveror}) are derived by (\ref{DeltaS:derivation}). 

\begin{figure}
	\centering
	\includegraphics[width=5cm]{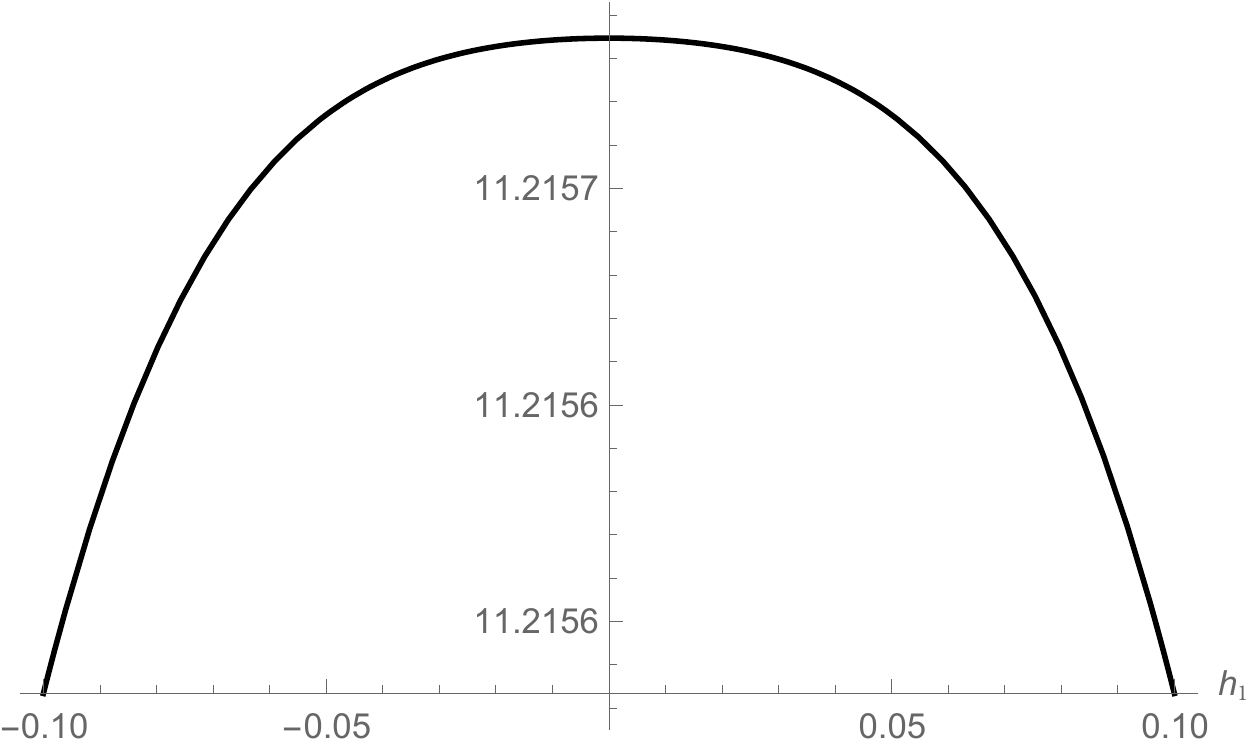} 
	\includegraphics[width=5cm]{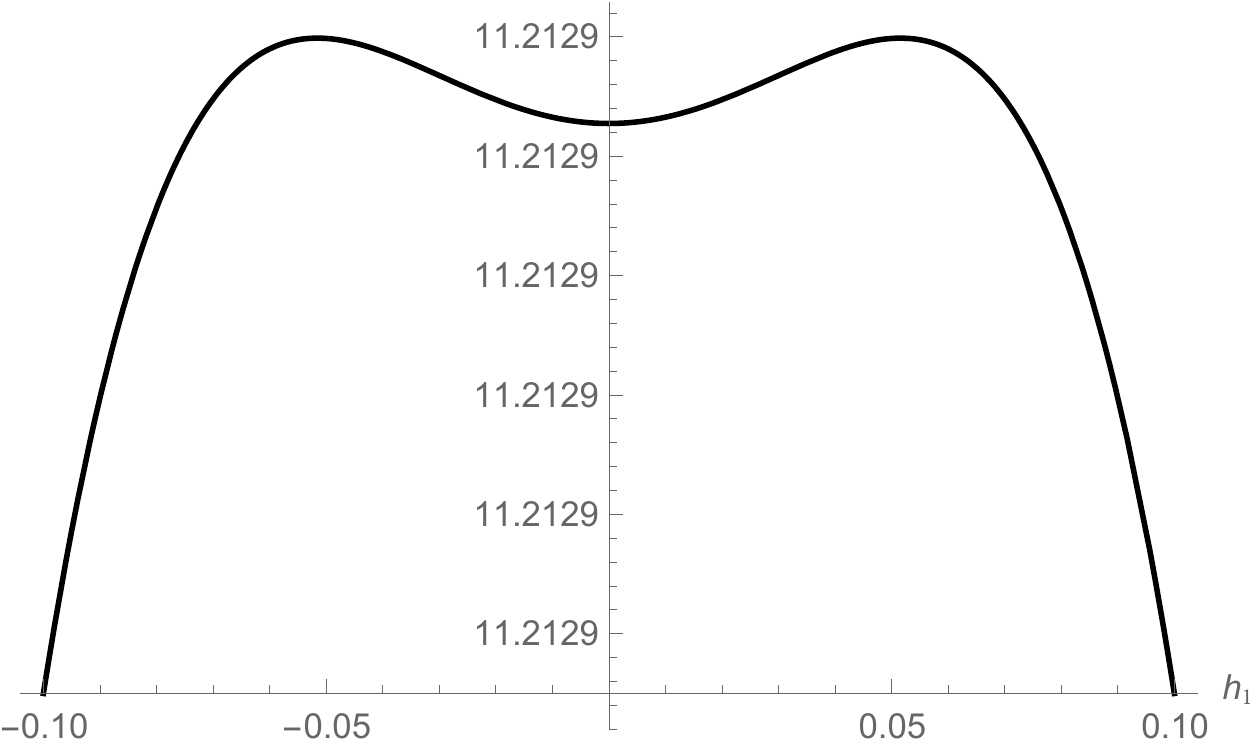} 
	\\
	(a) \hspace*{4.5cm} (b)
	\caption{
		An action functional $S(q+h_1 \psi^{(1)})$ for $\Delta N=-1$. 
		Bifurcation is supposed to be one side in $\xi>\xi_0$.
		(a) $\xi < \xi_0$.  
		(b) $\xi > \xi_0$.  
	}
	\label{DS(h)}
\end{figure}

%Based on $\Delta N$, (\ref{lambdaDeltaS}) and (\ref{DeltaS:behaveror}), 
On the basis of $\Delta N$, (\ref{lambdaDeltaS}) and (\ref{DeltaS:behaveror}), 
we can picture bifurcation 
through manifold of action functional $S$ 
in the subspace %of eigenfunctions 
of corresponding $\lambda$.
For example,  
suppose $\Delta N=-1$ and bifurcation is one side in $\xi>\xi_0$.
%When $\xi < \xi_0$ is increasing and approaching to $\xi_0$, 
% at the 
Thus, 
in the one dimensional subspace of $\lambda$, 
top of a local maximum in $S$ where $q$ locates for $\xi <\xi_0$, 
%in the left side of the bifurcation point $\xi_0$
shown in figure~\ref{DS(h)}~(a), 
will slightly cave in for $\xi > \xi_0$,  
which yield critical points for $q^b$ 
in the both sides of the cave, 
shown in figure~\ref{DS(h)}~(b). 
%since $\Delta S \ge 0$ for $\xi > \xi_0$ by (\ref{DeltaS:behaveror}). 
%and the bifurcating solution $q^b$ at a critical point with higher action 
%approaches to the $q$.
%For $\xi > \xi_0$ and $\xi$ continue increasing, 
%then $q$ locates at the top of a local maximum 
%and the $q^b$ with lower action value diverges. 
%
We confirmed that 
%sign of $\Delta S$ 
the inequalities (\ref{lambdaDeltaS}) and (\ref{DeltaS:behaveror}) 
corresponding to the picture of bifurcation as shown in figure \ref{DS(h)} 
hold %at all bifurcation points 
in our numerical calculations.

%Then 
%In order to compare number of bifurcated solutions with $\Delta N$,
We define an equivalent class, 
congruent class of bifurcated solutions $B$,
by regarding the bifurcated solutions with congruent orbits as equivalent, 
and denote the number of elements by $\#B$.
Then we define number of incongruent bifurcated solutions as 
\begin{equation}\label{NB:def}
	N_B=\lim_{\xi' \to \xi-0} \#B(\xi')+\lim_{\xi' \to \xi+0} \#B(\xi').
\end{equation}
In the following sections 
we will show %all points for $\Delta N \ne 0$ yield bifurcation and %further
\begin{equation}\label{N_B=DN}
	N_B = |\Delta N|
\end{equation}
holds for $\xi=\xi_0$ in our numerical calculations.
% (\ref{N_B=DN+1}) holds at bifurcation points. 

Note that 
at a bifurcation point the Morse index may not change, $\Delta N=0$,
if $\lambda=0$ but $d \lambda(\xi)/d\xi=0$. 
However for figure-eight choreographies in this paper, 
such point %with $\lambda=0$ and $\Delta N=0$ %are not isolated,  
never contribute to any bifurcation 
and belongs to $\lambda=0$ in the whole region of $\xi$ corresponding to the conservation laws. 
In other words, (\ref{N_B=DN}) holds not only for $\xi=\xi_0$ but also for all $\xi$. 

We call the region $\xi<\xi_0$ the left side of the bifurcation point $\xi_0$ and
$\xi_0<\xi$ the right side.

\section{Morse index and bifurcation for homogeneous system}
\label{sec:homo}

In this section we investigate the bifurcation of the figure-eight choreography 
by using $a$ as the parameter $\xi$ for
a system under homogeneous potential
\begin{equation}\label{key}
	U(q)=\sum_{b>c} u_a(|(q_{2 b-1}-q_{2 c-1},q_{2 b}-q_{2 c})|), \; 
	u_a(r)=-\frac{1}{r^a}, 
\end{equation}
for $a \ge 0$.
In table \ref{homo:tbl},   
$a$, $\Delta N(a) \ne 0$ and 
symmetry of the variated orbit $Q$ 
are tabulated.
The symbol $D$ means that the $Q$ is not choreographic, and 
the subscript $y$ indicates that the orbits are symmetric in the $y$ axis.
There is no other point with $\Delta N(a) \ne 0$ for $a \ge 0$ than 
tabulated in table \ref{homo:tbl}, $a=0.9966$ and $1.3424$. 

\begin{table}%-------------------------------------------------------
	\centering
	\caption{
		$\Delta N(a)$ and symmetry of $Q$ for $a \ge 0$.
	}
	\begin{tabular}{c c c}
		\hline
		$a$ & $\Delta N(a)$ & $Q$ \\
		\hline
		$0.9966$ & $-2$ & $D_y$ \\
		$1.3424$ & $-2$ & $D$ \\
		\hline
	\end{tabular}
	\label{homo:tbl}
\end{table}%-------------------------------------------------------

%There are two points, $a=0.9966, 1.3424$ at which $N(a)$ changes.
%We confirmed numerically that all these points are bifurcation points and there 
%is no point with    

\subsection{Bifurcation at $a=0.9966$}
\label{sec:simoH}

At $a=0.9966$, the Morse index $N(a)$ changes by $\Delta N=-2$ as shown in table \ref{homo:tbl}.
For $g=|\Delta N|=2$, the variated orbit $Q$ is written as
\begin{equation}\label{q+psi:D}
	Q=q + h(\cos\Theta \psi^{(1)}+ \sin\Theta \psi^{(2)})
\end{equation}
with $c_1=\cos\Theta$ and $c_2=\sin\Theta$.
The coefficients are, thus,  %$\cos \Theta$ and $\sin \Theta$ in (\ref{q+psi:D}) 
found as critical points in $\Theta$ by 
\begin{equation}\label{key}
	\frac{\partial S(Q)}{\partial \Theta}=0.
\end{equation}
%by minimizing $|q^b-Q|^2$ in \cite{fukuda2},
%
\begin{figure}
	\centering
	\includegraphics[width=5cm]{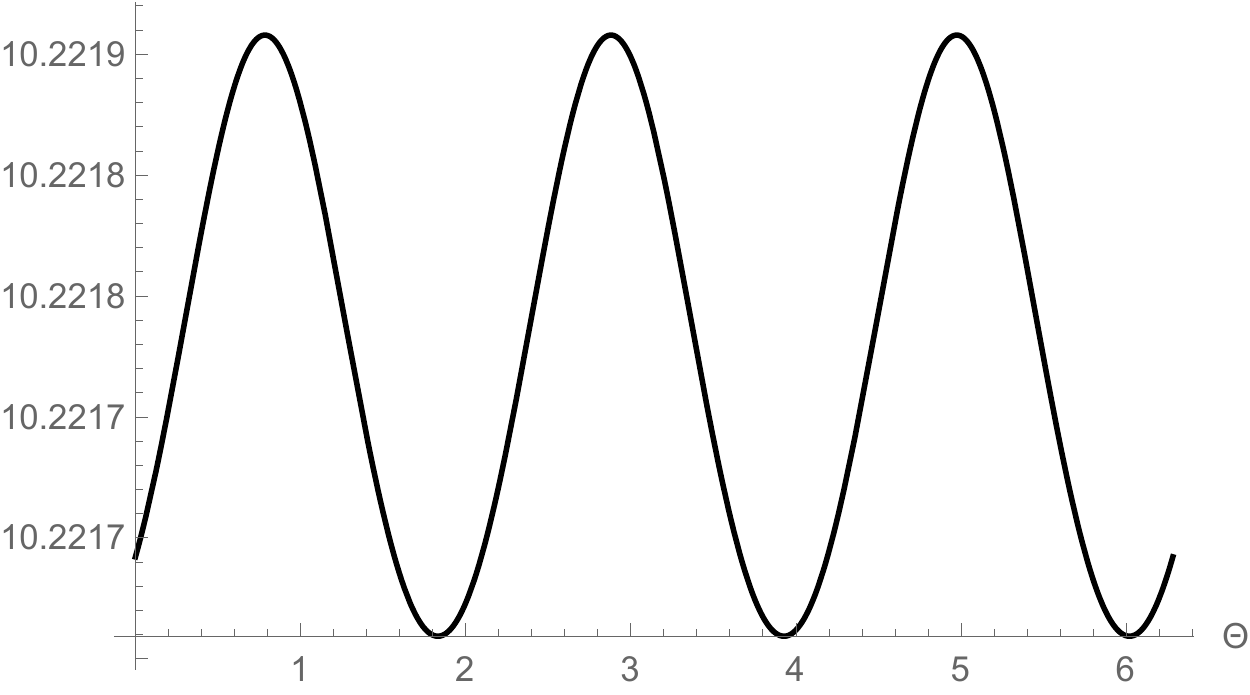} 
	\includegraphics[width=5cm]{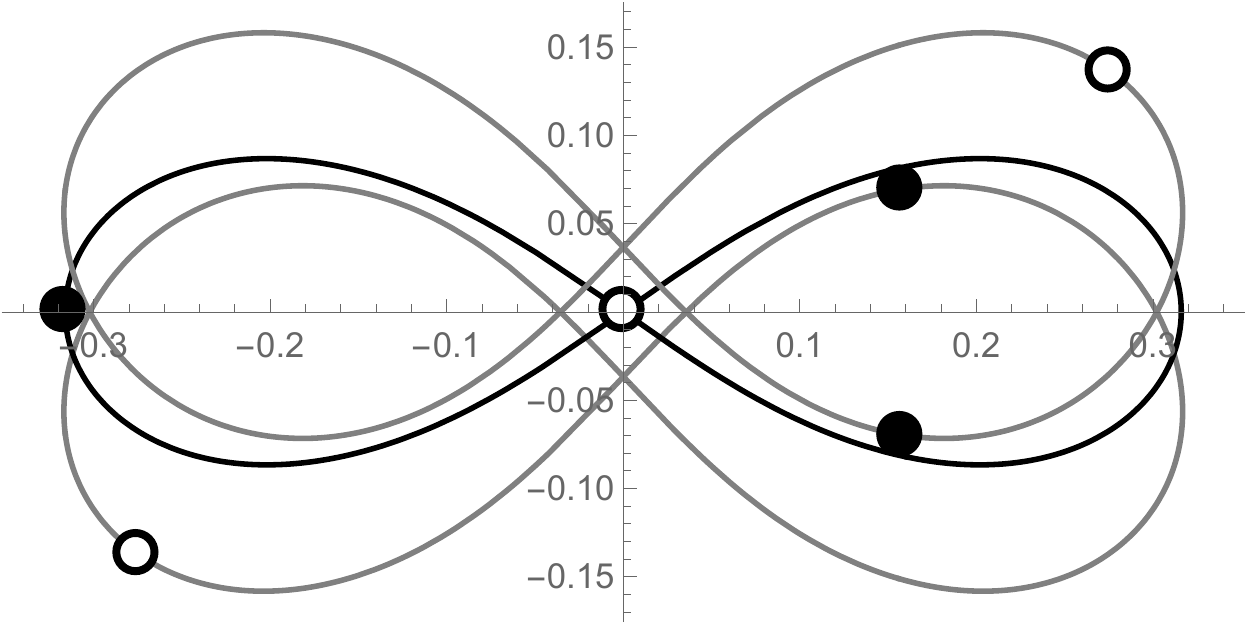} 
	\\
	(a) \hspace*{4.5cm} (b)
	\caption{
		Variated orbit $Q$ in (\ref{q+psi:D}) for 
		$a=1$ close to $a=0.9966$; 
		$T=1$, $h=0.08$, $\lambda=0.06881$. 
		(a) $S(Q)$ of $\Theta$.  
		(b) $Q$ with $D_{x y}$ symmetry %in (\ref{q+psi:D}) 
		%for $T=1$, $h=0.08$ and 
		for a local maximum at $\Theta=0.78343$ in $S(Q)$. 
		Note that the value of $\Theta$ does not have universal meaning since it depends on the choice of orthonormal basis, $\psi^{(1)}$ and $\psi^{(2)}$.
		Filled circles are isosceles triangle configuration at $t=0$ and open circles 
		Euler configuration at $t=T/4$. % for $D_{x y}$ orbits.			
	}
	\label{SofTheta}
\end{figure}

As shown in figure~\ref{SofTheta}~(a) at $h=0.08$, 
there are six critical points in $S(Q)$ of $\Theta$, 
three local maximums and three local minimums, 
which are independent of $h$. 
%
%Note that 
At all six critical points in $\Theta$, 
the variated orbits $Q$'s 
have the same symmetry higher than the $D_y$ indicated in table \ref{homo:tbl}.
They consist of 
three distinct orbits
symmetric in the $y$ axis and in the $x$ axis with exchange of two bodies; 
one orbit is symmetric itself but two collectively. 
Thus we denote orbits with this symmetry by $D_{x y}$. 
In figure~\ref{SofTheta}~(b), the variated orbit $Q$ at a local maximum is shown.
Black orbit is symmetric itself but two gray orbits collectively.
Consequently all bifurcating solutions 
from $a=0.9966$ 
will be searched within $D_{x y}$ symmetry.
Note that since it is numerically difficult to calculate the $Q$ just at the bifurcation point 
$a=0.9966$ and its symmetry does not depend on $a$, 
in figure~\ref{SofTheta} calculation for $a=1$ is shown.

Conditions for $q(t)$ to be $D_{x y}$, derived in \ref{A.H}, 
are 
that $q(t)$ takes an isosceles triangle configuration at $t=0$
shown by filled circles in figure~\ref{SofTheta}~(b), 
\begin{equation}\label{q.tri}
	q(0) = (x, y, -2 x, 0, x, -y), 
\end{equation}
\begin{equation}\label{p.tri}
	\dot{q}(0) = ( -\cos\theta, -\sin\theta, 0, 2 \sin\theta, \cos\theta, -\sin\theta) v, 
\end{equation}
with
\begin{equation}\label{H.theta}
	\theta=\tan^{-1} \frac{y}{3 x}, 
\end{equation}
by parameters $(x, y, v)$,
and 
an Euler configuration at $t=T/4$ 
shown by open circles in figure~\ref{SofTheta}~(b),  
\begin{equation}\label{H.Euler}
	( q_3, q_4, \dot{q}_1 \dot{q}_6 - \dot{q}_2 \dot{q}_5 )=0.
\end{equation}
For given period $T$,  
the three conditions (\ref{H.Euler}) determine the three parameters $(x, y, v)$.

In order to distinguish $D_{x y}$ from figure-eight choreography 
which are very close around a bifurcation point, 
we use the $y$ component of body 1 on the $x$ axis, 
\begin{equation}\label{key}
	d=q_2(t),  
\end{equation}
for $t \simeq T/12$ with $q_1(t)=0$ 
since $d$ is zero if and only if $D_{x y}$ is choreographic. 

Mu\~{n}oz-Almaraz \etal \cite{Vanderbauwhede_mail,Munoz-Almaraz2} first found the bifurcation 
at $a=0.9966$ using AUTO \cite{AUTO1, AUTO2},
and recently we re-found it using the equations (\ref{q.tri})--(\ref{H.Euler}) 
with Newton's method: 
three $D_{x y}$ solutions 
corresponding to the three local maximums in $S(Q)$ of $\Theta$ 
bifurcate in the right side of the bifurcation point, 
and three 
%which correspond 
to the local minimums
in the left side.

%Denoting position $\Theta$ of a  local maximum as $\Theta_3$, 
The six critical points in figure~\ref{SofTheta}~(a) 
are written by a position of local maximum $\Theta_3$ 
as $\Theta_3+2 j\pi/3+k\pi$, $j=0,1,2$, $k=0,1$. 
Thus the $Q$ for the six solutions bifurcating in the both sides %at $a=0.9966$  
are represented by $\Theta=\Theta_3+2 j\pi/3$
with a smooth increasing function $h$ of $a$.
Using the choreographic operator $\hat{C}$ defined by
%Figure-eight choreography is a motion satisfying 
%invariant under choreographic operator
\begin{equation}\label{C.def}
	\hat{C} f_i(t) = f_{i+2}(t-T/3) 
\end{equation}
since for the doubly degenerate eigenvalue \cite{fukuda2, fujiwara} 
\begin{equation}\label{Chat}
	\hat{C} (\cos\Theta \psi^{(1)}+\sin\Theta \psi^{(2)}) =
	\cos(\Theta+2\pi/3) \psi^{(1)}+\sin(\Theta+2\pi/3) \psi^{(2)}, 
\end{equation} 
the $Q$ %(\ref{q+psi:D}) 
for $D_{x y}$ is written as
\begin{equation}\label{qH}
	Q^{D_{x y}} = q + h \hat{C}^j (\cos\Theta_3 \psi^{(1)}+\sin\Theta_3 \psi^{(2)}),
	%= \hat{C}^j (q + h  (\cos\Theta_3 \psi^{(1)}+\sin\Theta_3 \psi^{(2)})), 
	\; 	j=0,1,2.
\end{equation}
%
%, and its value does not have any meanings. 

\begin{figure}
	\centering
	\includegraphics[width=5cm]{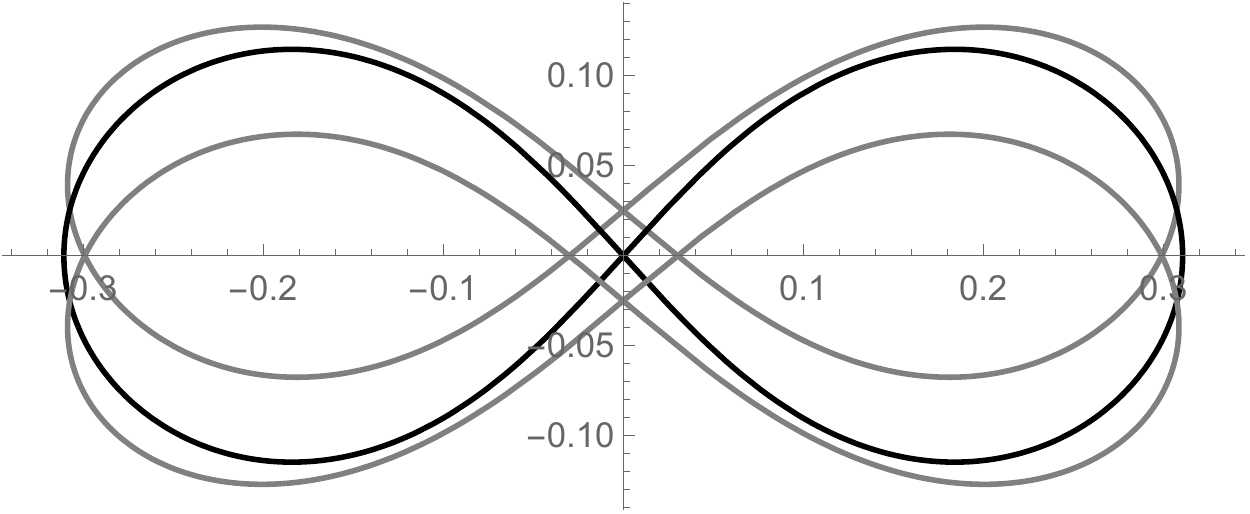} 
	\includegraphics[width=5cm]{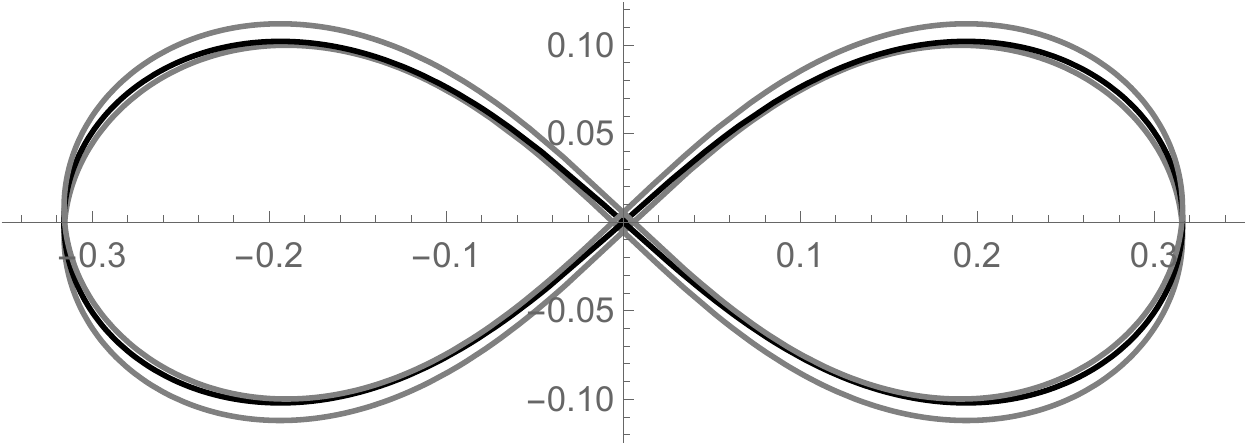} 
	\includegraphics[width=5cm]{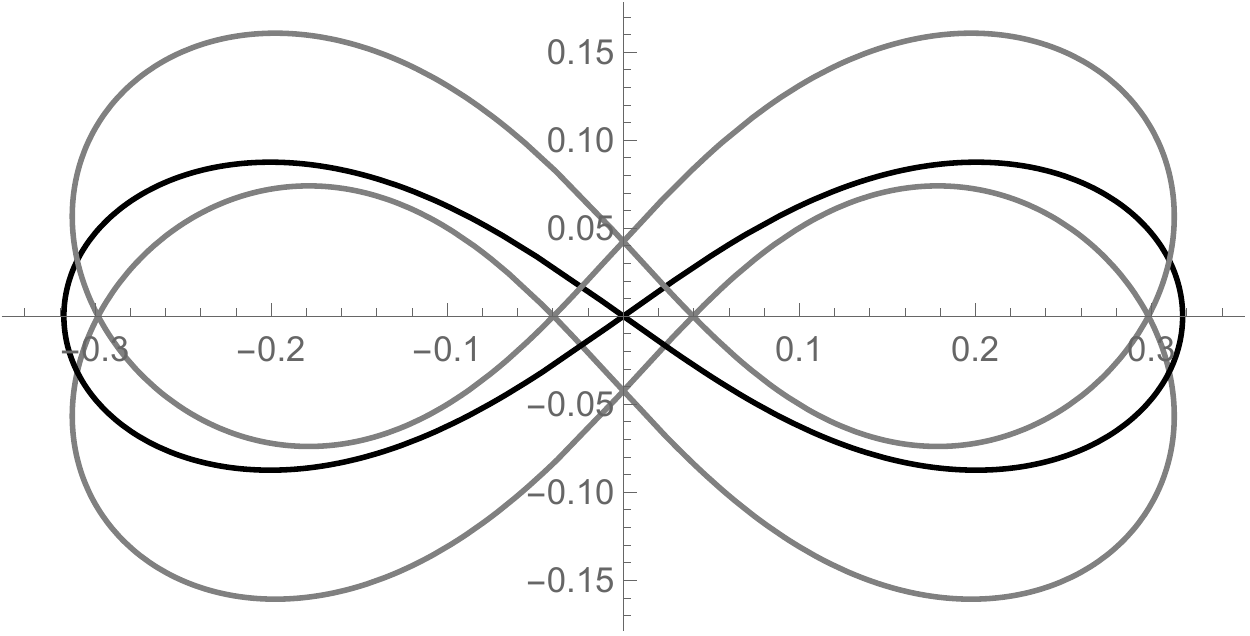} 
	\\
	(a) \hspace*{4.5cm} (b) \hspace*{4.5cm} (c)
	\caption{
		$D_{x y}$ solutions bifurcated from $a=0.9966$;   
		(a) $a=0.9766$, 
		(b) $a=1$,  and 
		(c) $a=1.0166$, for $T=1$.  
		Parameters $(x, y, v; d)$ for (a)--(c) are $(0.15533, 0.12092, 1.79372; 0.025027)$, 
		$(0.15800, 0.096588, 2.2321; -0.0054265)$ and $(0.15886, 0.072493, 2.58066; -0.042498)$,  respectively. 
	}
	\label{simoH}
\end{figure}

Since $\hat{C}^3=1$ and $\hat{C} q = q$, 
%\begin{equation}\label{q.choreo}
%	\hat{C} q(t) = q(t),
%\end{equation}
representation (\ref{qH}) for $j=0, 1, 2$ differ only
in cyclic permutation of bodies with time shift.
Thus their orbits are congruent and %for solutions bifurcating , 
the number of incongruent solutions $N_B$ is counted as $N_B=1+1=2$.  %at $a=0.9966$.
In figure~\ref{simoH} (a) and (c), 
one of three congruent $D_{x y}$ solutions 
in the both sides of bifurcation point, 
$a=0.9966 \pm 0.02$, %, $a=0.9766$ and $1.0166$,
are shown with parameters $(x,y,v)$ for initial condition and the index $d$.

As we showed in \cite{fukuda2}, 
%the Sim\'{o}'s H solution is written by the $Q$ in good agreement 
the Sim\'{o}'s H solution is in good agreement with the variated orbit $Q$ 
because it is the solution $D_{x y}$ 
bifurcating from $a=0.9966$ very close to $a=1$. 
Actually the solution $D_{x y}$ at $a=1$ shown 
in figure~\ref{simoH}~(b) coincides with the Sim\'{o}'s H solution.

\subsection{Bifurcation at $a=1.3424$}
\label{sec:D}
For $a \ge 0$, 
there is one more point changing the Morse index 
at $a=1.3424$ as shown in table \ref{homo:tbl}.
In this section we investigate this point in similar manner as in section \ref{sec:simoH}.

\begin{figure}
	\centering
	\includegraphics[width=5cm]{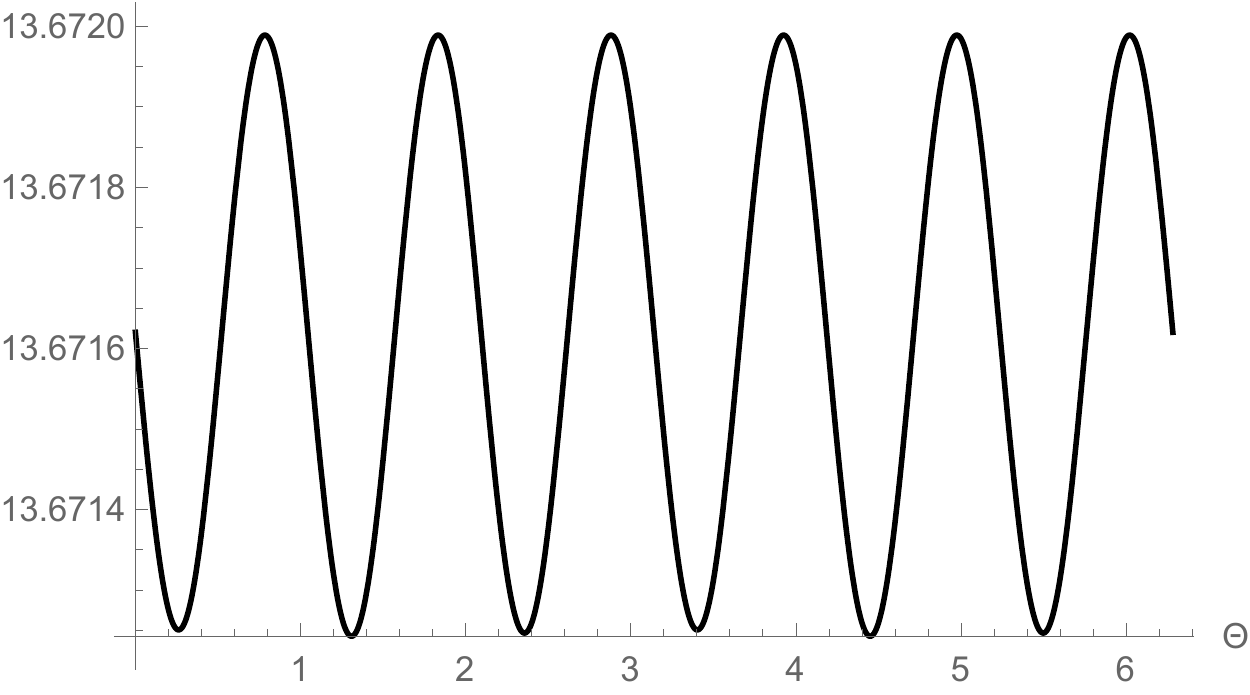} 
	\includegraphics[width=5cm]{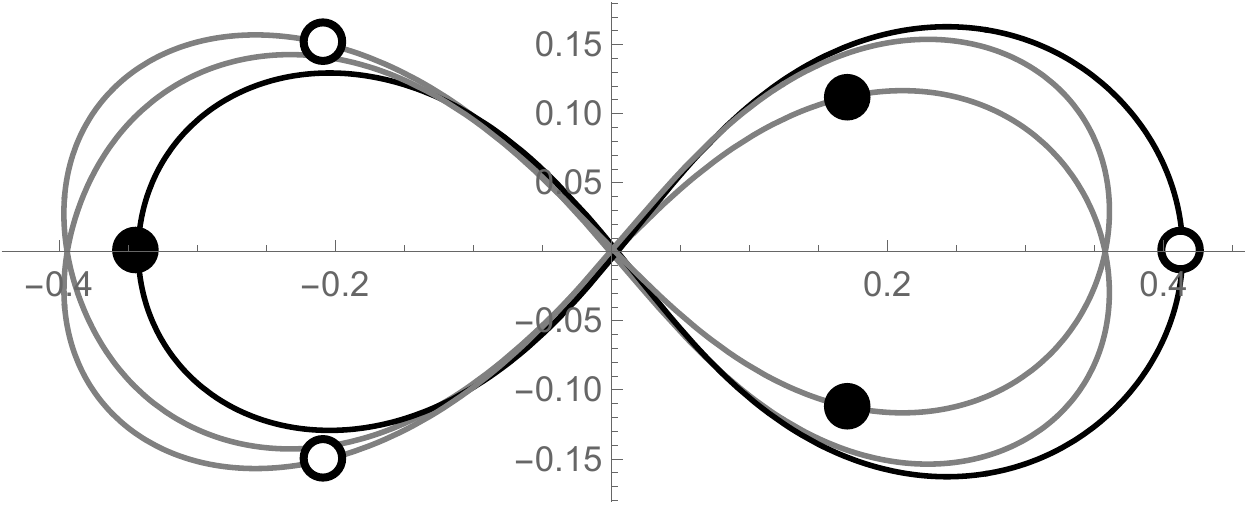} 
	\includegraphics[width=5cm]{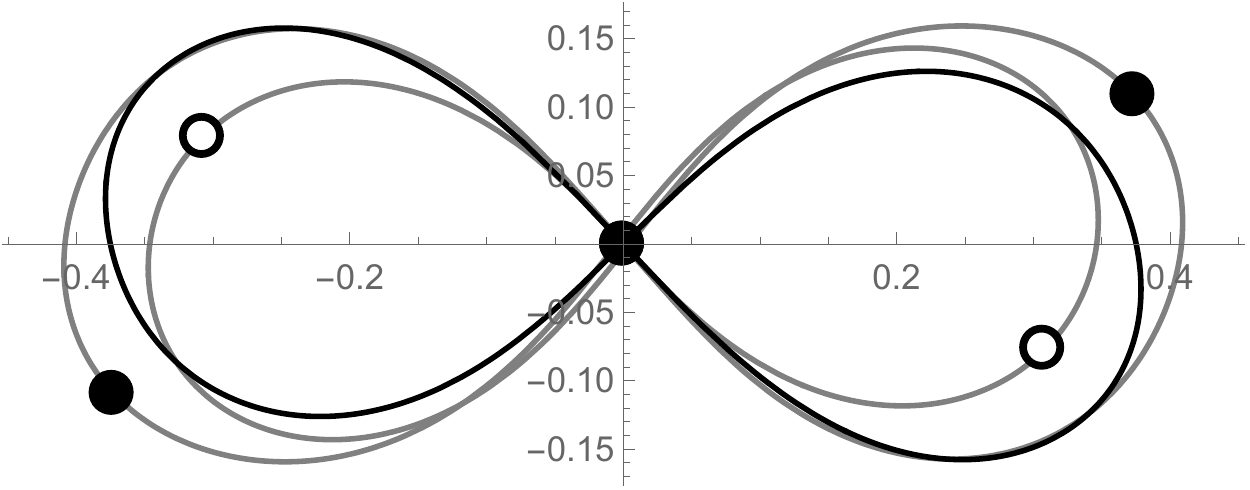} 
	\\
	(a) \hspace*{4.5cm} (b) \hspace*{4.5cm} (c)
	\caption{
		Variated orbit $Q$ %in (\ref{q+psi:D})  
		for $a=1.345$ close to $a=1.3424$; 
		$T=1$, $\lambda=0.02095$, $h=0.1$. 
		(a) $S(Q)$ of $\Theta$. 
		(b) $Q$ with $D_x$ symmetry %in (\ref{q+psi:D}) 
		%for $T=1$, $h=0.08$ and 
		for a local maximum at $\Theta=0.78448$. 
		Note that the $\Theta$ does not have universal meaning 
		since it depends on the choice of orthonormal basis, $\psi^{(1)}$ and $\psi^{(2)}$.
		Filled circles are isosceles triangle configuration at $t=0$ and open circles $t=T/2$. 
		(c) $Q$ with $D_2$ %in (\ref{q+psi:D}) 
		%for $T=1$, $h=0.08$ and 
		for a local minimum at $\Theta=0.78448+\pi/6$. 
		Filled circles are Euler configuration at $t=0$ and open circles at $t=T/2$.
	}
	\label{D.fig}
\end{figure}

The eigenvalue $\lambda$ which goes to zero at $a=1.3424$ is doubly degenerate,
$g=|\Delta N|=2$, and 
the variated orbit $Q$ in (\ref{q+psi:D}) are expected as bifurcating solutions.  
However its symmetry $D$ shown in table \ref{homo:tbl} is lower than for $D_{y}$ at $a=0.9966$, 
and its action $S(Q)$ as a function of $\Theta$ 
exhibits twelve critical points; 
six local maximums and six local minimums 
as shown in figure~\ref{D.fig}~(a).  
In figure~\ref{D.fig}, $S(Q)$ and $Q$ for $a=1.345$ 
are shown as a close point to $a=1.3424$ because of numerical convenience.  
%Note that the symmetry of $Q$'s at critical points are exchanged 
%if a point with $a<1.3424$ is chosen. 

At local maximums, the variated orbit $Q$ consists of three distinct orbits:
one orbit is symmetric itself in the $x$ axis and 
the other two are collectively,  
shown in figure~\ref{D.fig}~(b). %exchanged by the symmetry operations each other. 
On the other hand, 
at local minimums:
one orbit is symmetric itself at origin and 
the other two collectively, 
shown in figure~\ref{D.fig}~(c). %are exchanged by the symmetry operations each other. 
We denote former by $D_x$ and latter $D_2$.
Here the orbits of $D_x$ can have non zero total angular momentum $l$
%\begin{equation}\label{key}
%	J=\sum_b \bm{r}_b \times \dot{\bm{r}}_b
%\end{equation}
since 
sum of signed area of the three orbits can be non zero 
whereas it is zero for solutions $D_2$, $D_{x y}$ and the figure-eight choreography
because of two fold symmetry at origin.

Conditions for $q(t)$ to be $D_x$, derived in \ref{A.Dx}, are
that $q(t)$ takes an isosceles triangle configuration 
shown by filled circles in figure~\ref{D.fig}~(b), 
(\ref{q.tri}) and (\ref{p.tri}) 
with 
\begin{equation}\label{D.tri.l}
	\theta = \tan^{-1} \frac{y}{3 x}-\tan^{-1} \frac{l}{2 v\sqrt{9x^2+y^2}}, \label{Dy.tri}
\end{equation}
at $t=0$ 
by parameters $(x, y, v, l)$, 
and opposite isosceles triangle configuration at $t=T/2$ 
shown by open circles in figure~\ref{D.fig}~(b), 
\begin{equation}\label{D.con}
	( q_4, \dot{q}_3, q_1-q_5, \dot{q}_2-\dot{q}_6 )=0. %\; \dot{q}_4>0.
\end{equation}
For given period $T$,  
four conditions (\ref{D.con}) determine four parameters $(x, y, v, l)$.  
An index to distinguish $D_x$ from figure-eight choreography is 
\begin{equation}\label{key}
	(d, l) \ne 0, 
\end{equation}
since $D_x$ with $l=0$ is the $D_{x y}$ solution.

Conditions for $q(t)$ to be $D_2$, derived in \ref{A.D2}, are
that $q(t)$ takes an Euler configuration at $t=0$ 
shown by filled circles in figure~\ref{D.fig}~(c), 
\begin{equation}\label{q.Euler}
	q(0)=(x,0,-x,0,0,0), 
\end{equation}
\begin{equation}\label{p.Euler}
	\dot{q}(0)=(u,v,u,v,-2 u,-2 v) 
\end{equation}
by parameters $(x, u, v)$, 
and another Euler configuration at $t=T/2$ 
shown by open circles in figure~\ref{D.fig}~(c), 
\begin{equation}\label{D2.con}
	( q_5, \, q_6, \dot{q}_1\dot{q}_4-\dot{q}_2\dot{q}_3)=0. %\; \dot{q}_3(t)>0.
\end{equation}
%See figure~\ref{D.fig} (b).
For given period $T$, 
three conditions (\ref{D.con}) determine three parameters $(x, u, v)$. 
%are  obtained by three dimensional Newton's method. 
%
An index to distinguish $D_2$ from figure-eight choreography is 
\begin{equation}\label{key}
	(d, \Delta I) \ne 0, 
\end{equation}
where $\Delta I=I(T/2)-I(0)$ and $I(t)=|q(t)|^2=\sum_i |q_i(t)|^2$, 
since $D_2$ with $\Delta I=0$ is the $D_{x y}$ solution.

Using AUTO \cite{AUTO1, AUTO2}, 
Mu\~{n}oz-Almaraz \etal  
found the solution $D_x$ bifurcated at $a=1.3424$ in 2005
together with $D_{x y}$ at $a=0.9966$ \cite{Vanderbauwhede_mail,Munoz-Almaraz2}. 
Knowing the existence of $D_x$, we recently re-found $D_x$ and $D_2$ 
using the above equations with Newton's method: 
six $D_x$ and six $D_2$ solutions bifurcate 
in the right side of bifurcation point. %, $a>1.3424$.

\begin{figure}
	\centering
	\includegraphics[width=5cm]{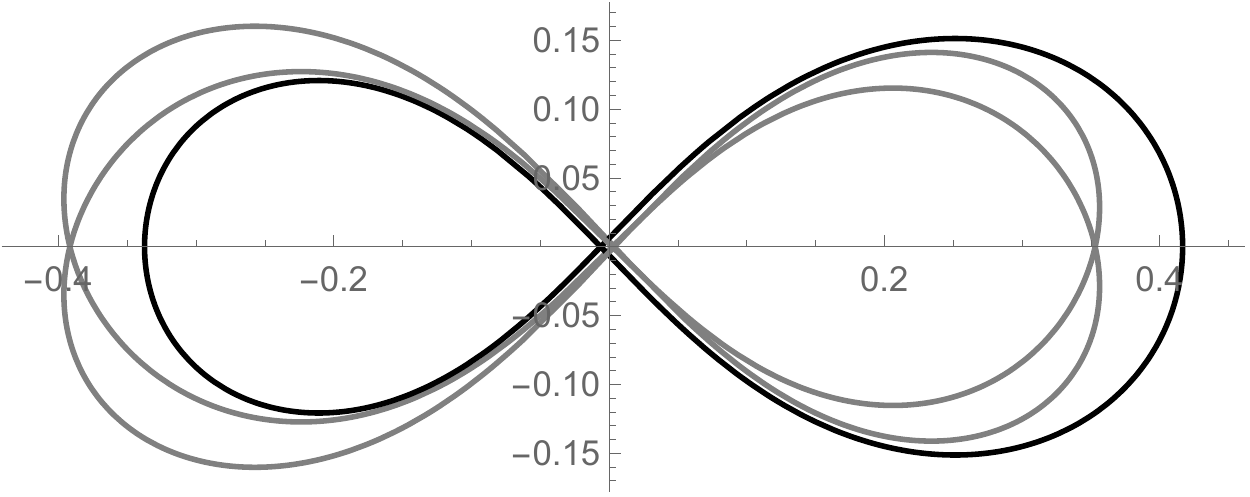} 
	\includegraphics[width=5cm]{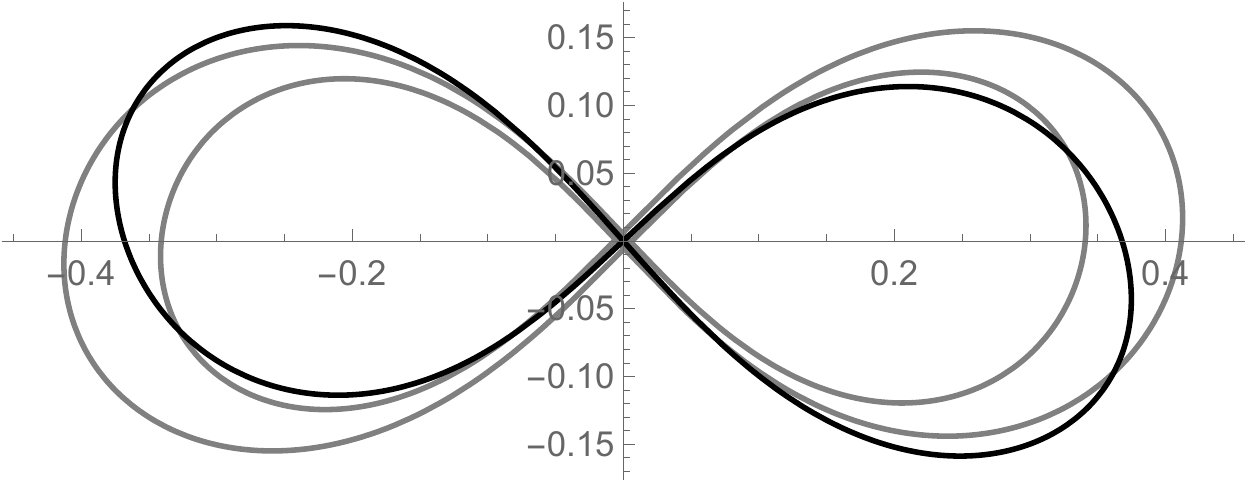} 
	\\
	(a) \hspace*{4.5cm} (b) %\hspace*{4.5cm} (c)
	\caption{
		(a) $D_x$ and (b) $D_2$ solutions at $a=1.3425$ with $T=1$ bifurcated from $a=1.3424$. 
		$D_2$ is rotated by $\theta_2$ in (\ref{theta2}). 
		%Black curve is symmetric itself but the two gray curves collectively.
		Parameters: (a) $(x, y, v, l)=(0.16872, 0.11111, 2.8874, 0.000035433)$, 
		(b) $(x, u, v; \Delta I)=(0.39375, 0.57707, 1.1697; -0.11584)$. 
	}
	\label{DxD2}
\end{figure}

The six $D_x$ solutions are written by the $Q$ %in (\ref{q+psi:D}) 
%with $h>0$ and 
with $\Theta$ at local maximums, 
and the six $D_2$ at local minimums.
The $\Theta$'s at local maximums are written 
by a position of local maximum $\Theta_6$ % by symmetry.
as $\Theta_6+2 j\pi/6$ 
with integer $j$, 
and local minimums  as $\Theta_6+(2 j+1)\pi/6$. 
Thus, by (\ref{Chat}), the $Q$ for $D_x$ and $D_2$ solutions are represented by 
\begin{equation}\label{Dx.Q}
  Q^{D_x} = q \pm h \hat{C}^j (\cos\Theta_6 \psi^{(1)}+\sin\Theta_6 \psi^{(2)}), 
\end{equation}
and 
\begin{equation}\label{D2.Q}
  Q^{D_2} = q \pm h \hat{C}^j (\cos(\Theta_6+\pi/6) \psi^{(1)}+\sin(\Theta_6+\pi/6) \psi^{(2)}), 
\end{equation}
respectively, with $j=0,1,2$ and $a>1.3424$.
In (\ref{Dx.Q}) and (\ref{D2.Q}), 
the sign in front of $h$ effects inversion of orbits in the $y$ axis.  
Then the orbits of six $D_x$ solutions
are all congruent, 
in direct isometry or mirror inversion, 
and the orbits of six $D_2$ solutions are so. 
Thus the number of incongruent bifurcating solutions $N_B$ is counted as $N_B=0+2=2$.
%are also congruent
%
In figure~\ref{DxD2}, one of six congruent $D_x$ and $D_2$ solutions at $a=1.3425$ 
are shown with parameters for initial conditions and with $\Delta I$ for $D_2$, 
where $D_2$ is rotated by 
\begin{equation}\label{theta2}
	\theta_2=-\frac{1}{2} \tan^{-1}\frac{q_4(T/2)}{q_3(T/2)}, 
\end{equation}
to make the $x$ axis bisector of two Euler configurations.

%We call the bifurcation discussed in this section $D_x/D_2$ type, according to the symbol 
%for the symmetry of the corresponding variated function in \cite{fukuda2}.

\section{Morse index and bifurcation for LJ system}
\label{sec:LJ}

In this section, 
we discuss the bifurcation of the $\alpha$ solution by using $T$ as the parameter $\xi$ 
for the system under LJ-type potential
\begin{equation}\label{key}
	U(q)=\sum_{b>c} u^{LJ}(|(q_{2 b-1}-q_{2 c-1},q_{2 b}-q_{2 c})|).
\end{equation}
The $\alpha$ solution bifurcates at $T=T_{\min}=14.479$, 
thus there exists no $\alpha$ solution for $T<T_{\min}$ and 
two $\alpha$ solutions for $T>T_{\min}$.
One branch $\alpha_-$ from $T=T_{\min}$ of $\alpha$ solution 
tends to the figure-eight choreography under homogeneous potential
with $a=6$ for $T \to \infty$,
and the other branch $\alpha_+$ 
gourd-shaped for $T \to \infty$ \cite{fukuda2}.

In table \ref{LJ:tbl}, 
$T$, $\Delta N(T) \ne 0$ and symmetry of the variated orbit $Q$ 
for $\alpha_+$ and $\alpha_-$ are tabulated.
Symbols, $C_x$, $C_y$, $C_{x y}$ and $C_2$ mean 
that the variated orbit $Q$ is choreographic and 
is symmetric in the $x$ axis, in the $y$ axis, in both the $x$ and the $y$ axes, and at origin, 
respectively.
Note that 
though the symbols $C_x$ and $C_{x y}$ were written as $C$ and $C_e$ in \cite{fukuda2}, respectively, 
we redefined them 
since $C$ and $C_e$ have the same symmetry 
as $D_x$ and $D_{x y}$,  
defined in section \ref{sec:D}, respectively. 
There is no other point with $\Delta N(T) \ne 0$ for $\alpha$ solution  
than seven points tabulated in table \ref{LJ:tbl}.

\begin{table}%-------------------------------------------------------
	\centering
	\caption{
		$\Delta N(T)$ and symmetry of $Q$ for LJ $\alpha_\pm$. 
		Symbols, $C_x$, $C_y$, $C_{x y}$ and $C_2$ mean 
		that the $Q$ is symmetric in the $x$ axis, in the $y$ axis, in both the $x$ and the $y$ axes and at origin, 
		respectively.
	} 
	\begin{tabular}{c c c|c c c}
		\multicolumn{3}{c}{$\alpha_-$} & \multicolumn{3}{c}{$\alpha_+$} \\
		\hline
		$T$ & $\Delta N(T)$ & $Q$ & $T$ & $\Delta N(T)$ & $Q$ \\
		\hline
		$14.479$ &          & $C_{x y}$ & $16.111$ & $2$ & $D$  \\
		$14.595$ & $-1$ & $C_x$ & $16.878$ & $2$ & $D_y$ \\
		$14.836$ & $-2$ & $D_y$ & $17.132$ & $1$ & $C_y$ \\
		$14.861$ & $-2$ & $D$ & $18.615$ & $1$ & $C_2$ \\
		\hline
	\end{tabular}
	\label{LJ:tbl}
\end{table}%-------------------------------------------------------

\begin{figure}
	\centering
	\includegraphics[width=6cm]{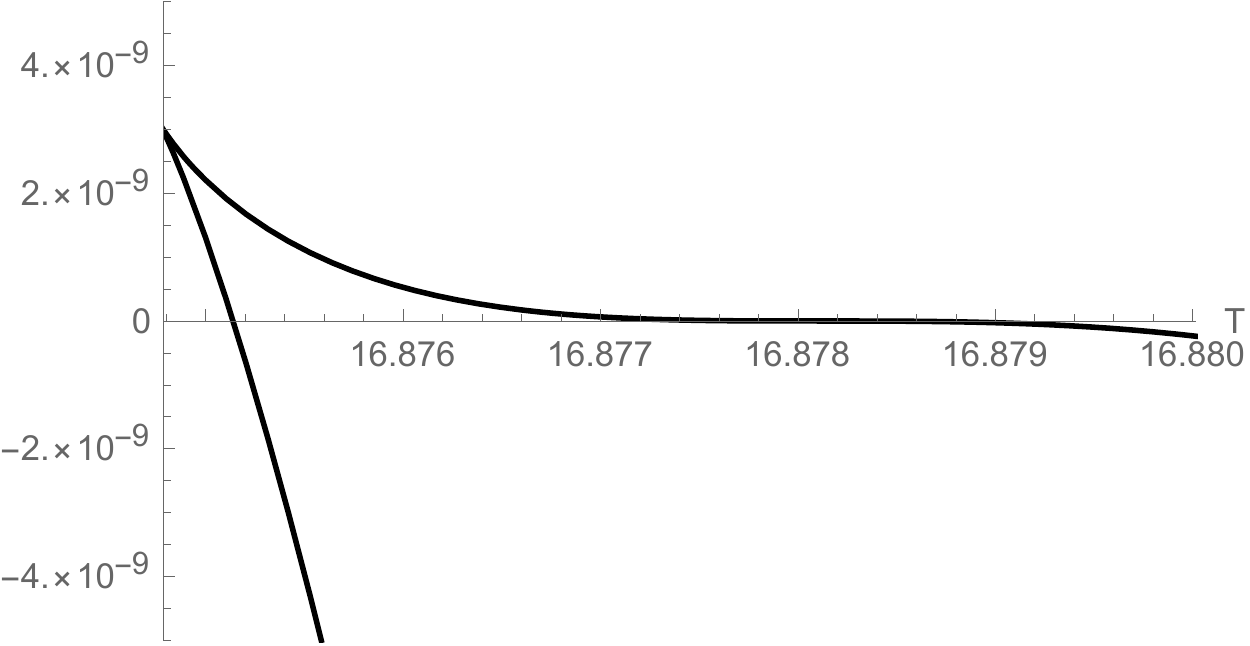}
	\includegraphics[width=4.5cm]{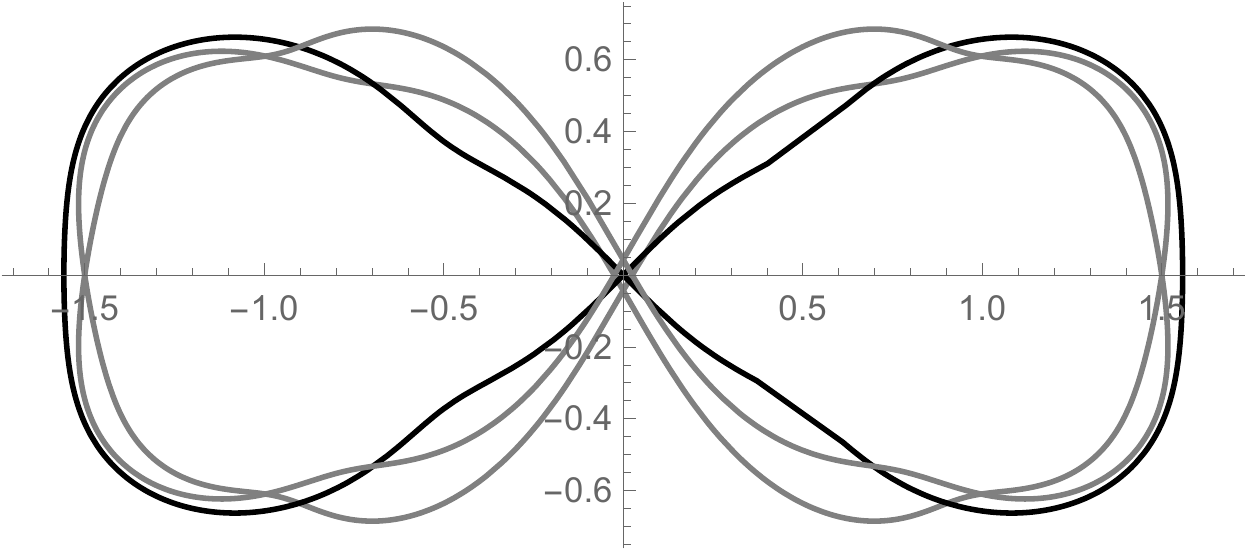}
	\includegraphics[width=4.5cm]{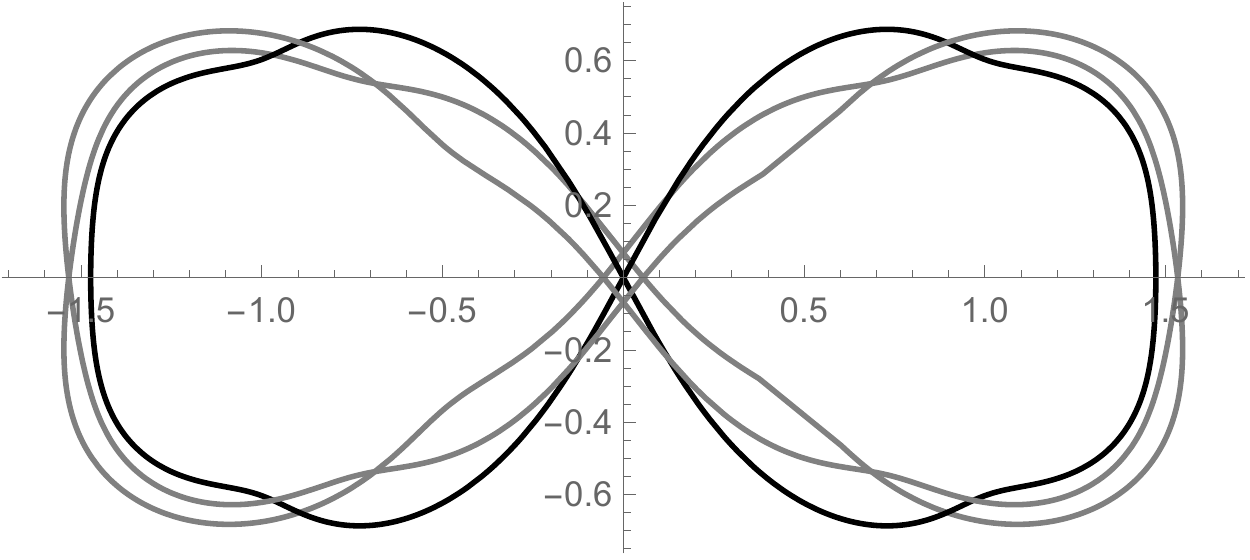}
	\\
	\hspace*{1cm} (a)　\hspace*{4.5cm} (b) \hspace*{3.7cm} (c)
	\caption{
		Bifurcation of $\alpha_+$ at $T=16.878$ yielding $D_{x y}$ solutions. 
		(a) $\Delta S(T)$, 		
		(b) $D_{x y}$ for $T=20$ from the right side of bifurcation point, and 
		(c) from the left side.
		Parameters $(x, y, v; d)$; (b) $(0.77903, 0.54721, 0.59844; -0.042907)$, 
		(c) $(0.73650, 0.54718, 0.59824; 0.068048)$.		
	}
	\label{Halpha+}
\end{figure}

\begin{figure}
	\centering
	\includegraphics[width=6cm]{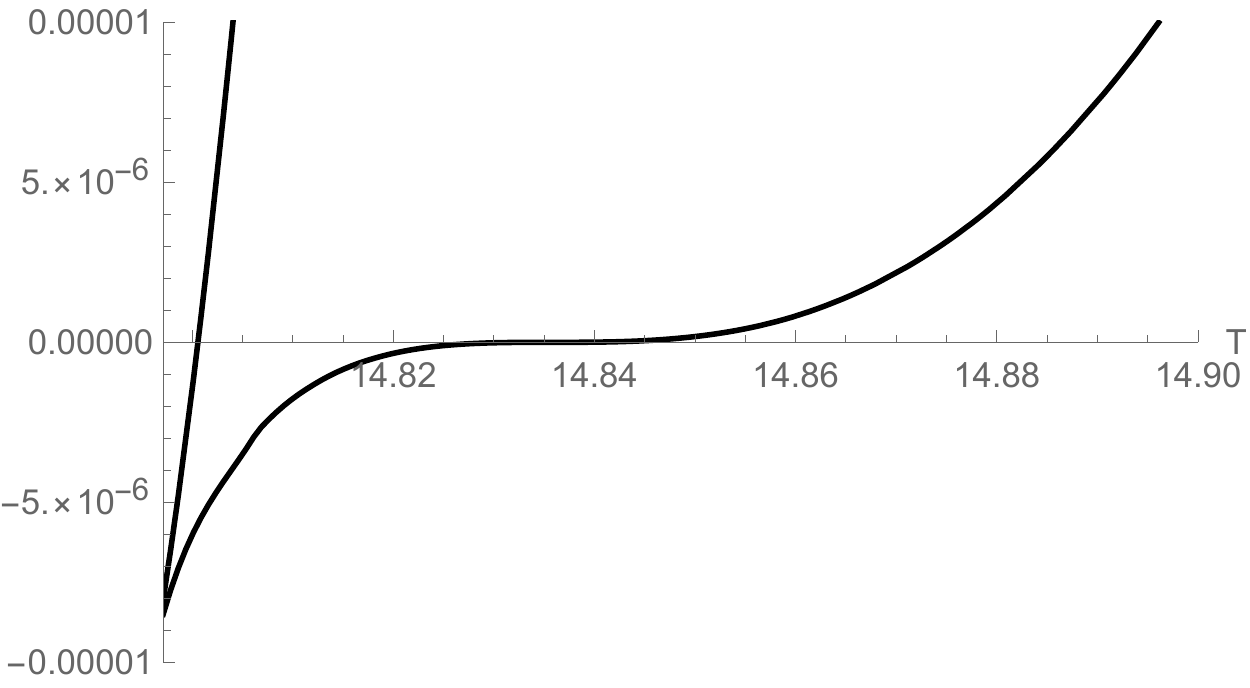}
	\includegraphics[width=4.5cm]{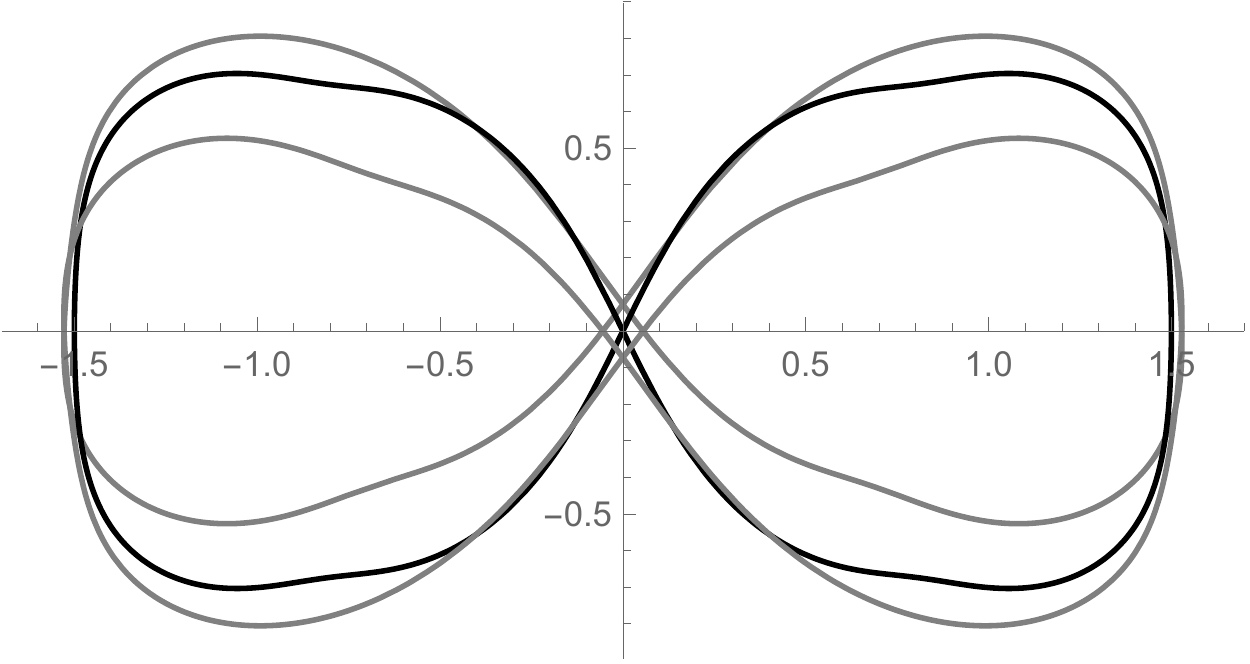}
	\includegraphics[width=4.5cm]{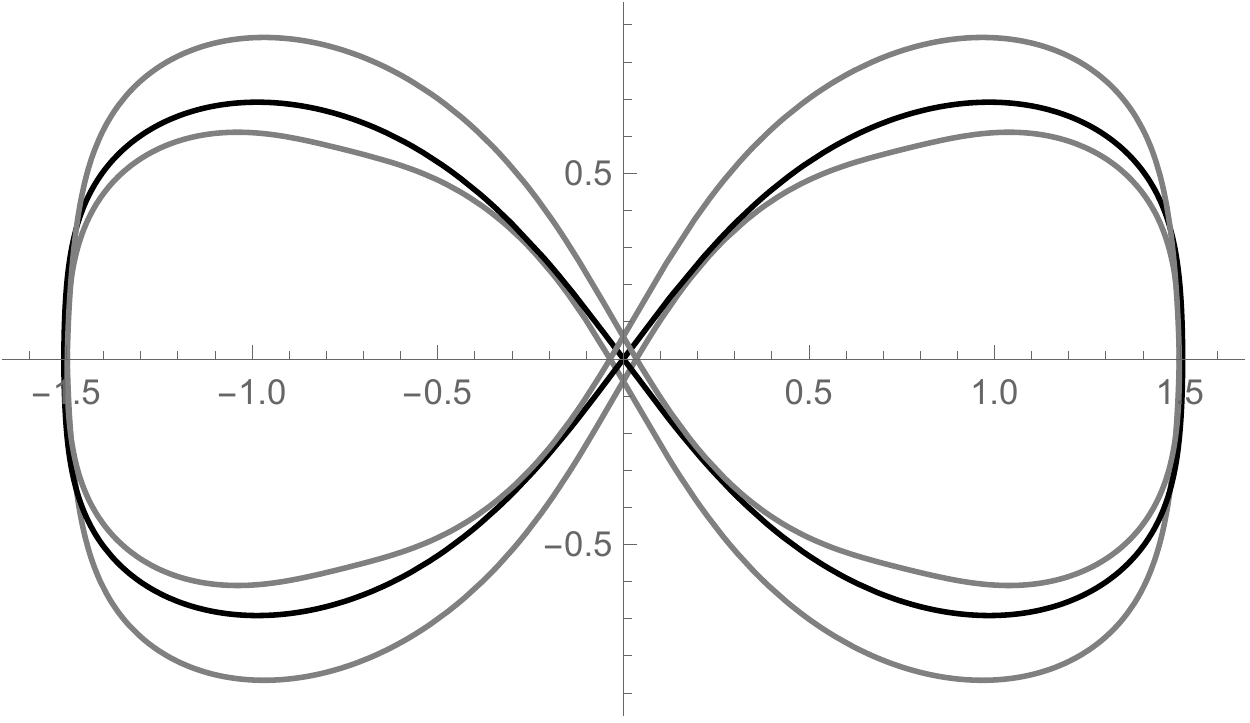}
	\\
	\hspace*{1cm} (a)　\hspace*{4.5cm} (b) \hspace*{3.7cm} (c)
	\caption{
		Bifurcation of $\alpha_-$ at $T=14.836$ yielding $D_{x y}$ solutions. 
		(a) $\Delta S(T)$, 
		(b) $D_{x y}$ for $T=20$ from the right side of bifurcation point, and 
		(c) from the left side.
		Parameters $(x,y,v; d)$: (b) $(0.74968, 0.76413, 0.45966; 0.072840)$,
		(c) $(0.75349, 0.56442, 0.64295; -0.059918)$.
	}
	\label{Halpha-}
\end{figure}

\subsection{Bifurcation yielding $D_{x y}$, $D_x$ and $D_2$ solutions}
\label{sec:LJH}

The points indicated by $D_{y}$ in table \ref{LJ:tbl}, 
$T=16.878$ for $\alpha_+$ and $T=14.836$ for $\alpha_-$,  
yield $D_{x y}$ solutions represented by (\ref{qH}) 
as in the section \ref{sec:simoH}.
The bifurcations are the both sides and 
the number of incongruent bifurcating solutions $N_B$'s are both two.

%There are bifurcations yielding $D_{x y}$ solutions, 
%discussed in the section \ref{sec:simoH} for homogeneous system, 
%at $T=16.878$ for $\alpha_+$ and $T=14.836$ for $\alpha_-$.
%As in the section \ref{sec:simoH}, 
%their corresponding variated orbits $Q$'s have $D_y$ symmetry shown in table \ref{LJ:tbl}, 
%and they are represented by (\ref{qH}). 
%The bifurcations are the both sides and 
%the number of incongruent bifurcating solutions $N_B$'s are both two.

In figure~\ref{Halpha+}~(a), 
$\Delta S(T)$ for bifurcation from $\alpha_+$ at $T=16.878$ is plotted. 
Though bifurcation is both sides, 
the bifurcating solution also bifurcate soon at $T=16.875$.  
Thus there exist two bifurcated solutions for $T>16.875$. 
For $16.875<T<16.878$ both solutions 
are bifurcated from left side of bifurcation point but 
for $T>16.878$ one from right side and the other from left side.
In figure~\ref{Halpha+}~(b) and (c), 
the two bifurcated solutions for $T=20$ 
from both sides %of $T=16.878$ 
are shown with parameters $(x,y,v)$ and $d$.
In figure~\ref{Halpha-}, 
bifurcation of $D_{x y}$ solution from $\alpha_-$ at $T=14.836$ 
is shown as figure~\ref{Halpha+} for $\alpha_+$.

\begin{figure}
	\centering
	\includegraphics[width=6cm]{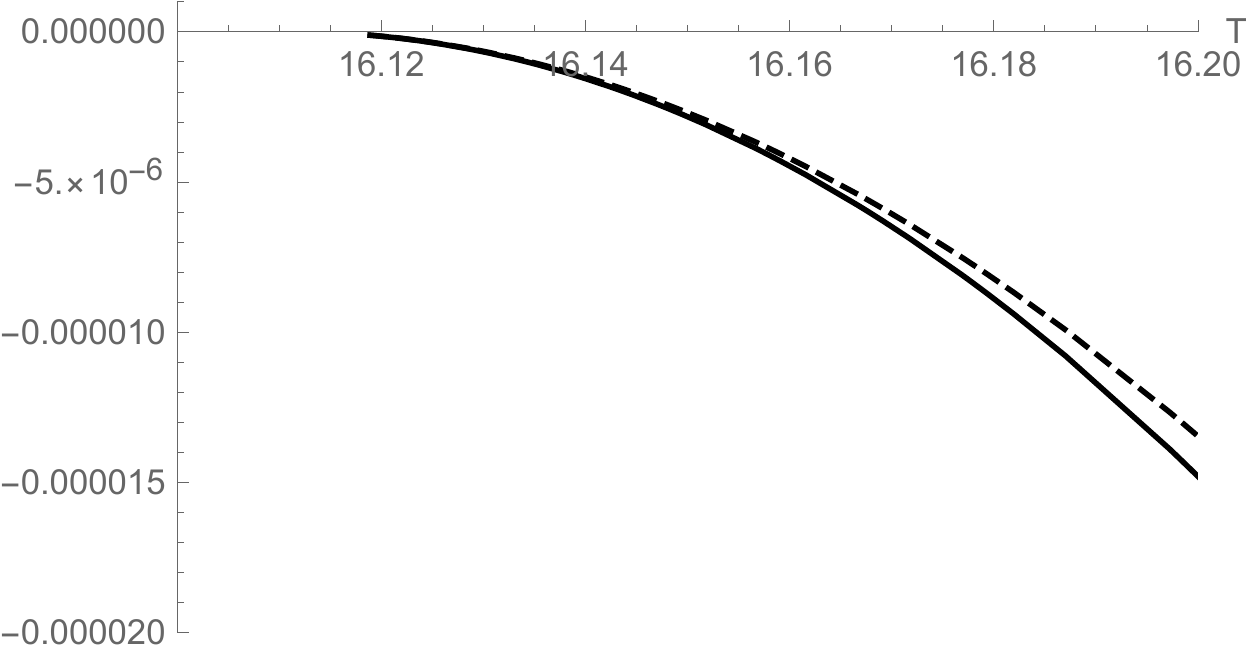}
	\includegraphics[width=4.5cm]{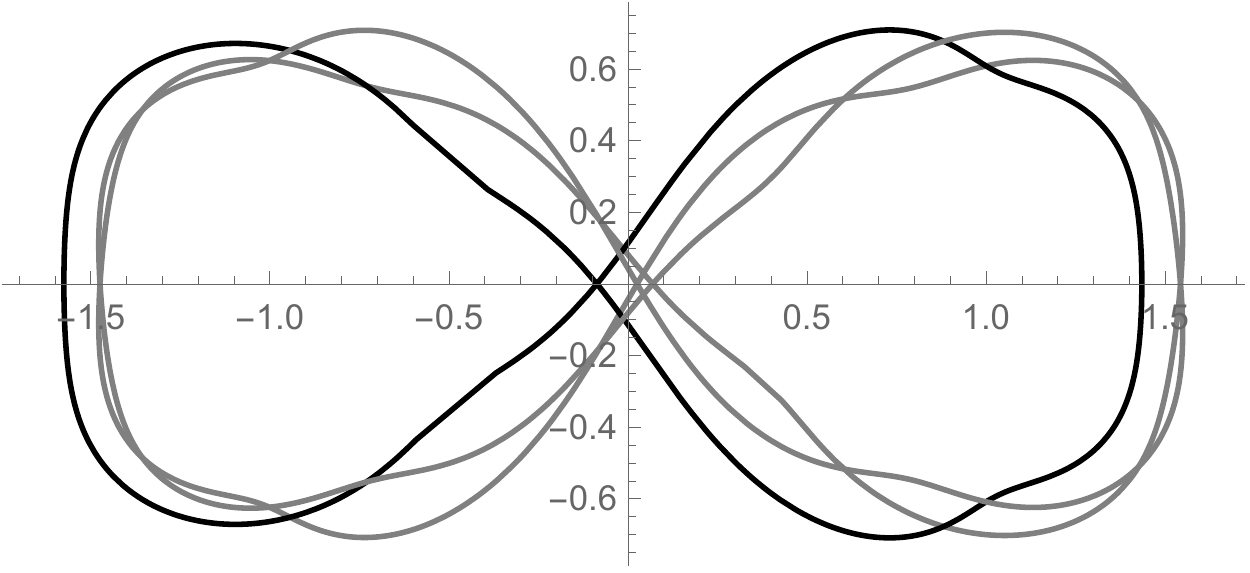}
	\includegraphics[width=4.5cm]{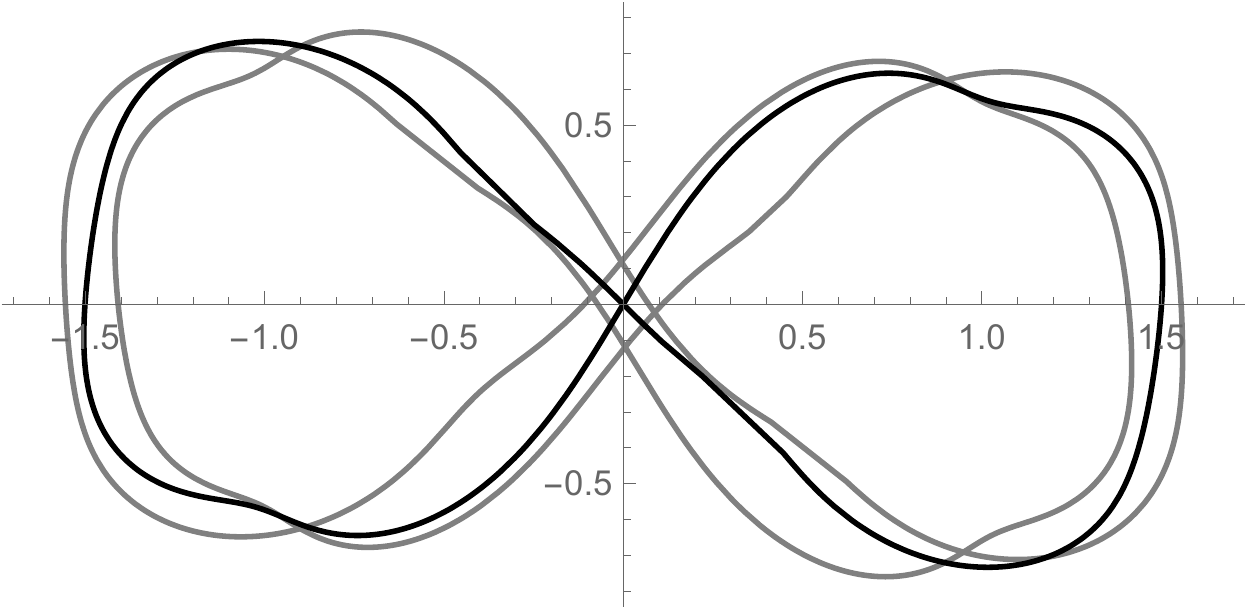}
	\\
	\hspace*{1cm} (a)　\hspace*{4.5cm} (b) \hspace*{3.7cm} (c)
	\caption{
		Bifurcation of $\alpha_+$ at $T=16.111$ yielding $D_x$ and $D_2$ solutions.
		(a) $\Delta S(T)$ for $D_x$ solution (full curve) and for $D_2$ (dashed curve). 		
		(b) $D_x$ solution and (c) $D_2$ for $T=20$. 
		$D_2$ is rotated by $\theta_2$ in (\ref{theta2}).
		Parameters; 
		(b) $(x,y,v,l)=(0.78752, 0.54620, 0.59854, 0.0028781)$,
		(c) $(x,u,v; \Delta I)=(1.3362, 0.16921, 0.34182, 1.3573)$.
	}
	\label{Dalpha+}
\end{figure}

\begin{figure}
	\centering
	\includegraphics[width=6cm]{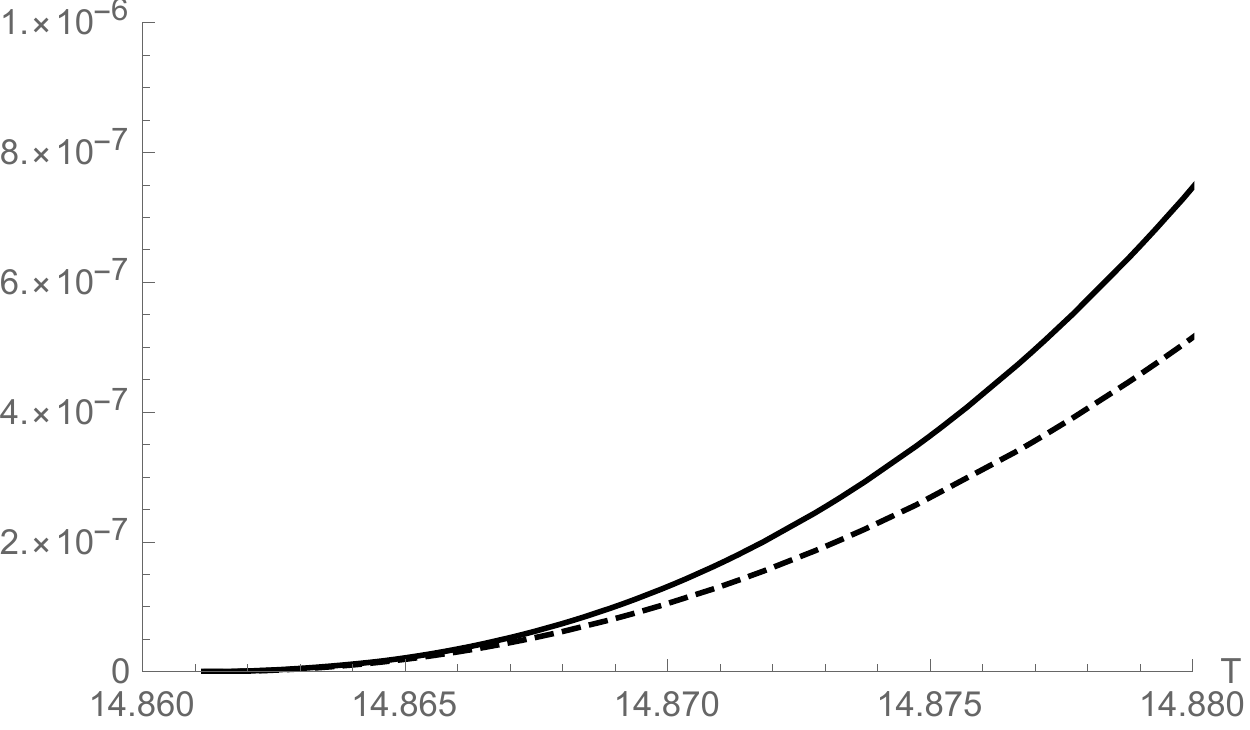}
	\includegraphics[width=4.5cm]{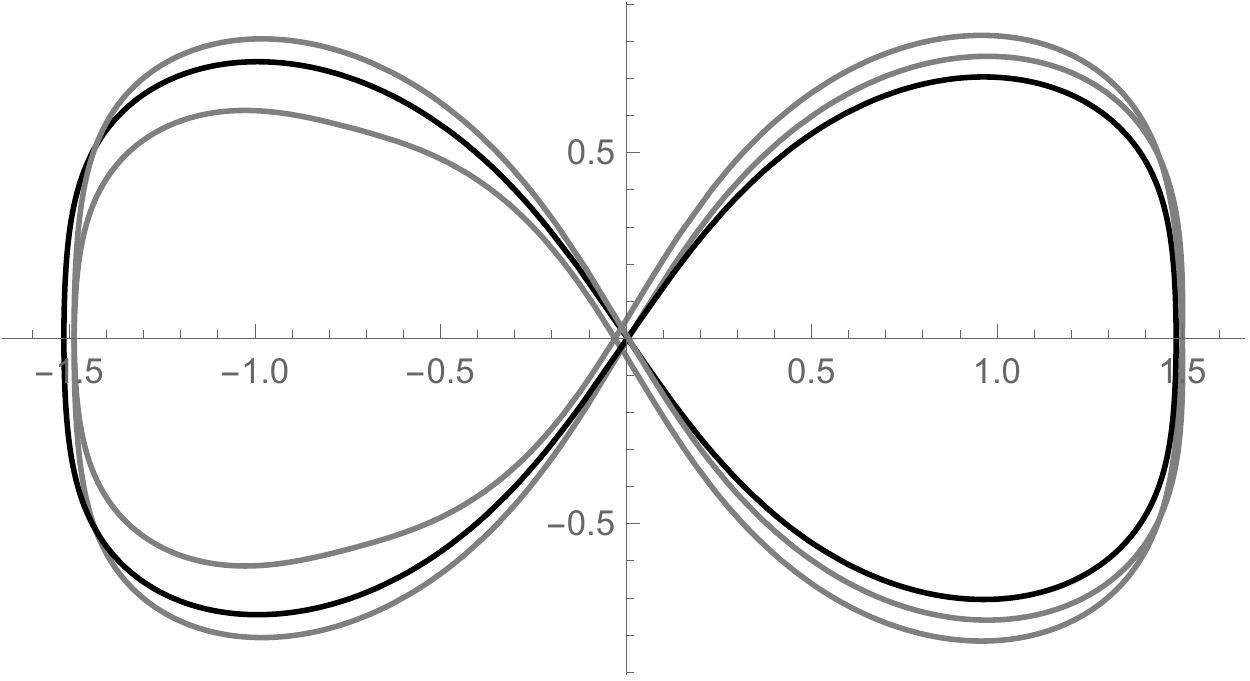}
	\includegraphics[width=4.5cm]{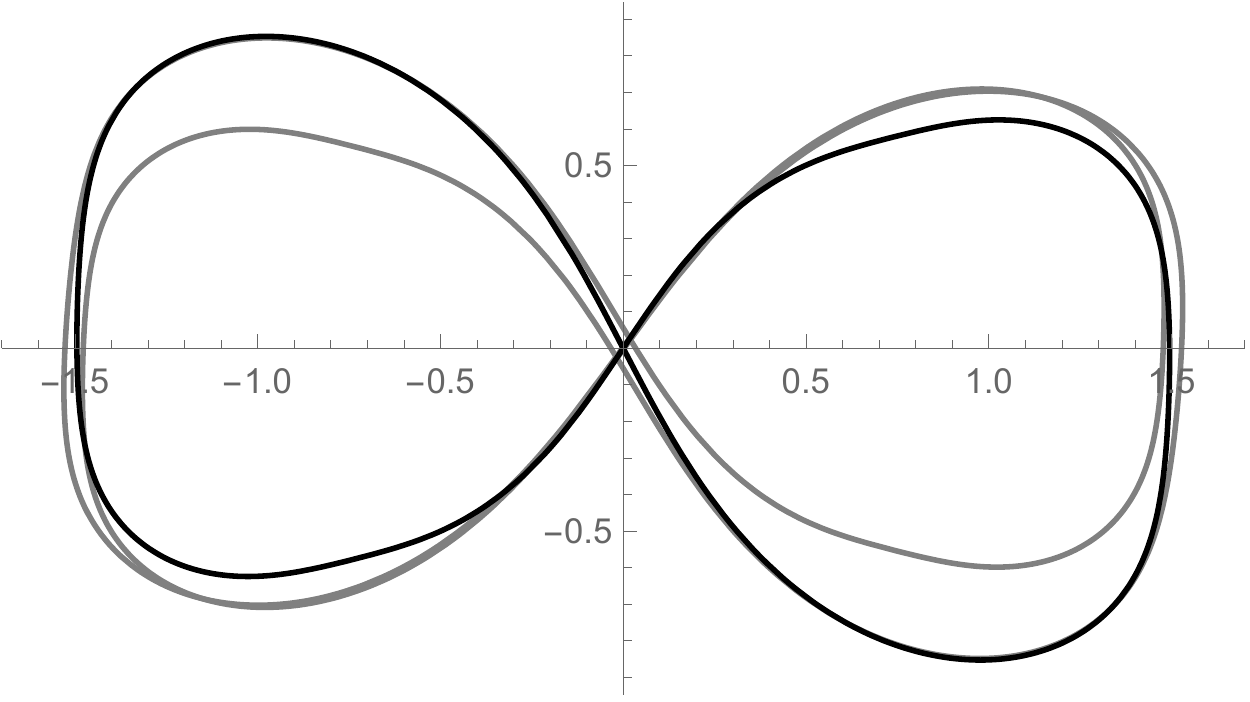}
	\\
	\hspace*{1cm} (a)　\hspace*{4.5cm} (b) \hspace*{3.7cm} (c)
	\caption{
		Bifurcation from $\alpha_-$ at $T=14.861$ yielding $D_x$ and $D_2$ solutions. 
		(a) $\Delta S(T)$ for $D_x$ solution (full curve) and for $D_2$ (dashed curve). 		
		(b) $D_x$ solution and (c) $D_2$ for $T=20$. 
		$D_2$ is rotated by $\theta_2$ in (\ref{theta2}).
		Parameters; 
		(b) $(x, y, v, l)=(0.75744, 0.72602, 0.52093, 0.027042)$, 
		(c) $(x, u, v; \Delta I)=(1.5333, 0.061120, 0.33256; -0.52266)$
	}
	\label{Dalpha-}
\end{figure}

The points indicated by $D$ in table \ref{LJ:tbl}, 
$T=16.111$ for $\alpha_+$ and $T=14.861$ for $\alpha_-$, 
yield $D_x$ and $D_2$ solutions represented by (\ref{Dx.Q}) and (\ref{D2.Q}), respectively, 
as in the section \ref{sec:D}. 
The bifurcations are one side in the right side and 
the number of incongruent bifurcating solutions $N_B$'s are both two. 
In figures~\ref{Dalpha+} and \ref{Dalpha-}, $\Delta S(T)$ and the bifurcated solutions for $T=20$ 
are shown with parameters. 

%There are bifurcations yielding $D_x$ and $D_2$ solutions, discussed in section \ref{sec:D}, 
%at $T=16.111$ for $\alpha_+$ and $T=14.861$ for $\alpha_-$.
%As in the section \ref{sec:D}, 
%their corresponding variated orbits $Q$'s have $D$ symmetry shown in table \ref{LJ:tbl}, 
%they are represented by (\ref{Dx.Q}) and (\ref{D2.Q}), respectively. 
%The bifurcations are one side in the right side and 
%the number of incongruent bifurcating solutions $N_B$'s are both two.

\subsection{Choreographic bifurcation}
\label{sec:LJC}

The points with $|\Delta N|=1$ in table \ref{LJ:tbl}, 
$T=18.615$ and $T=17.132$ for $\alpha_+$, and $T=14.595$ for $\alpha_-$, 
bifurcate choreographic solutions 
since the variated orbit $Q$ for $g=|\Delta N|=1$, 
\begin{equation}\label{q+psi:C}
	Q=q + h \psi^{(1)}, 
\end{equation}
is choreographic, $\hat{C}Q=Q$ \cite{fukuda2, fujiwara}. 

%We show that at three points, 
%$T=18.615$ and $T=17.132$ for $\alpha_+$, and $T=14.595$ for $\alpha_-$,
%choreographic solutions bifurcate. 
%At these points,  
%the Morse index $N(T)$ changes by one as shown in table \ref{LJ:tbl}, 
%and corresponding eigenvalue $\lambda$ is non degenerate.
%Thus the variated orbit $Q$ for $g=|\Delta N|=1$,
%\begin{equation}\label{q+psi:C}
%	Q=q + h \psi^{(1)}, 
%\end{equation}
%is choreographic, $\hat{C}Q=Q$ \cite{fukuda2, fujiwara}. 
%%
%In figure \ref{Cx.fig}, corresponding three variated orbits $Q$'s are shown 
%for $T$ not so close to the point with $|\Delta N|=1$ 
%since the symmetry of $Q$ is unchanged throughout $T$ \cite{fukuda2}.

\begin{figure}
	\centering
	\includegraphics[width=4.5cm]{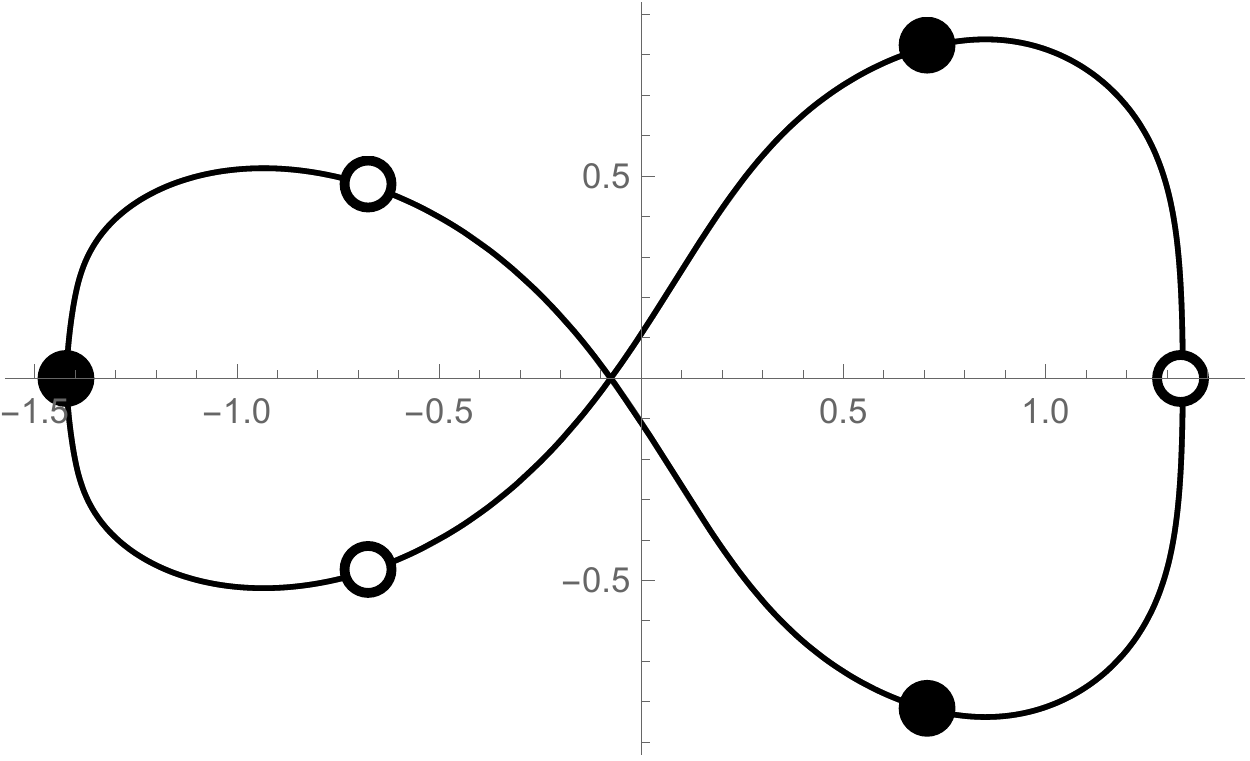}
	\includegraphics[width=4.5cm]{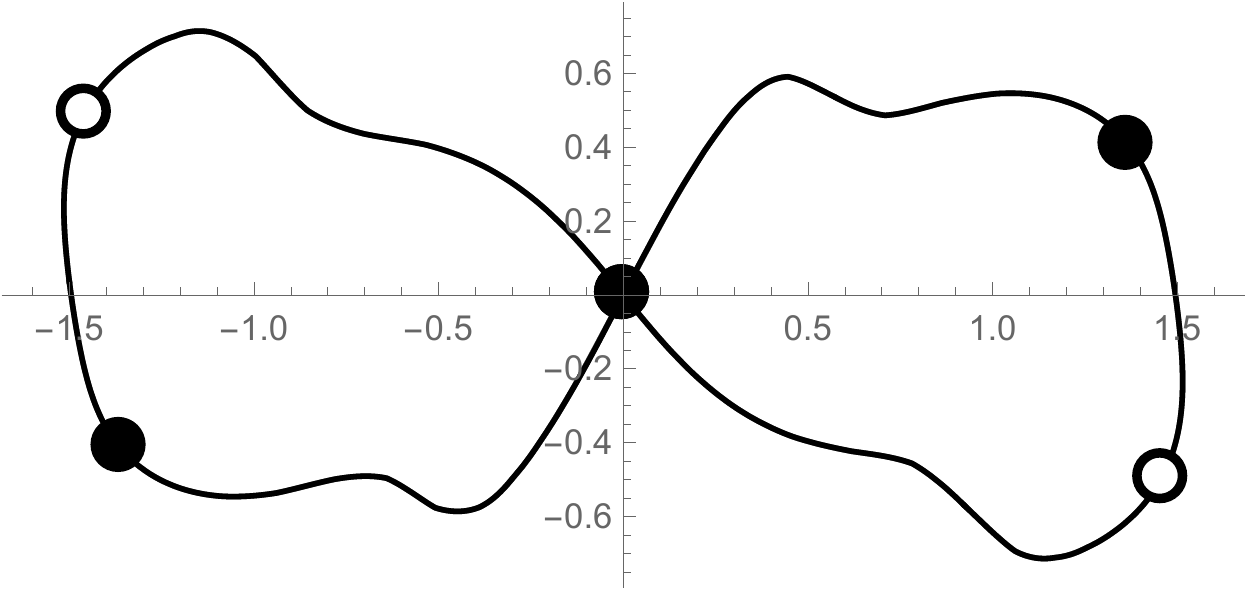}
	\includegraphics[width=4.5cm]{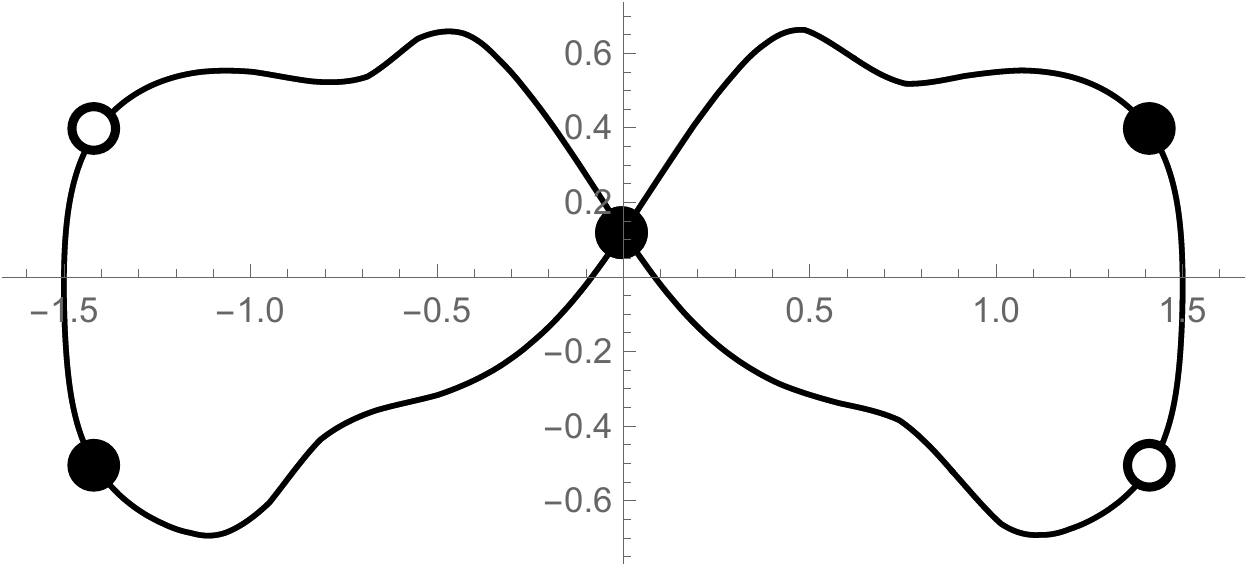}
	\\
	(a) \hspace*{4cm} (b) \hspace*{4cm} (c)
	\caption{
		Variated orbit $Q$ for LJ $\alpha$ solution. 
		(a) $C_x$ in $\alpha_-$; $T=15.215$, $h=1.5$ and $\lambda=0.43026$. 
		Filled circles are isosceles triangle configuration at $t=0$ and 
		open circles at $t=T/6$. 
		(b) $C_2$ in $\alpha_+$; $T=18.337$, $h=0.8$ and $\lambda=0.0027783$. 
		Filled circles are Euler configurations at $t=0$ and open circles at $t=T/6$. 
		(c) $C_y$ in $\alpha_+$; $T=18.337$, $h=0.8$ and $\lambda=-0.023907$. 
		Filled circles are configuration where a body on the $y$ axis at $t=0$ and 
		open circles at $t=T/6$.
	}
	\label{Cx.fig}
\end{figure}

\begin{figure}
	\centering
	\includegraphics[width=4.5cm]{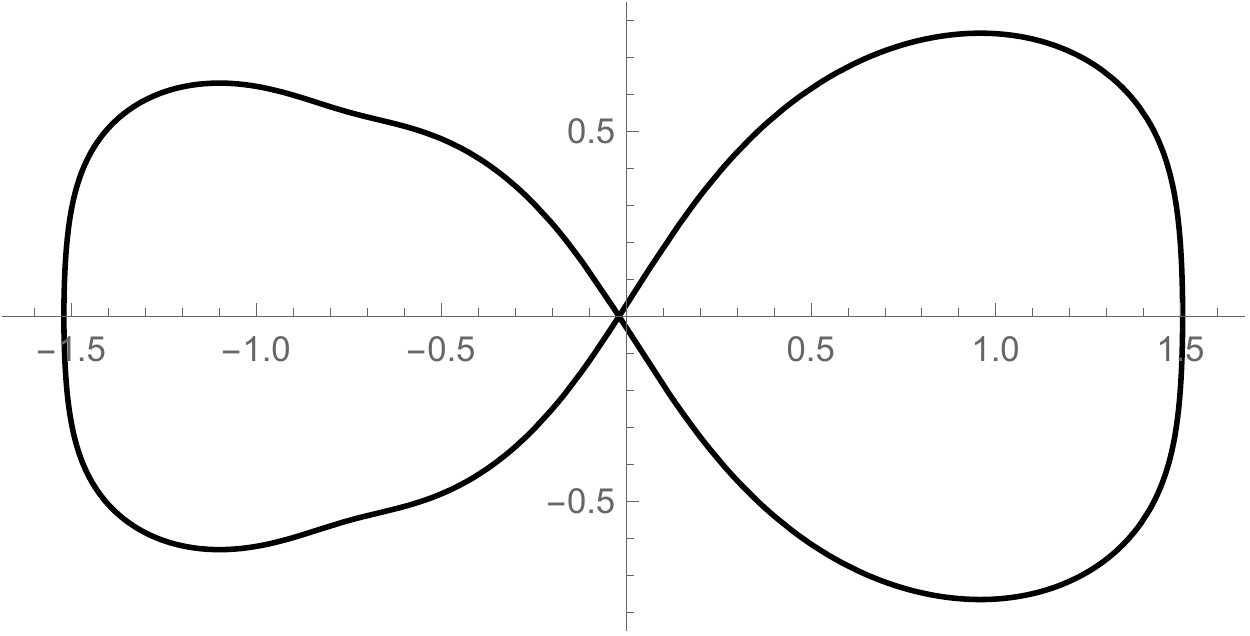}
	\includegraphics[width=4.5cm]{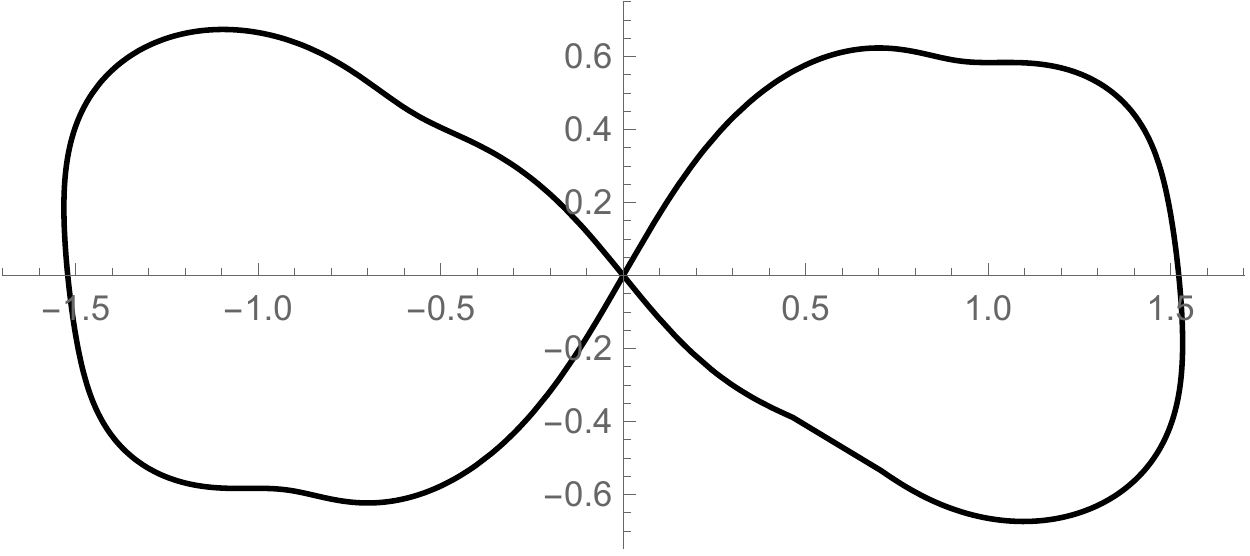}
	\includegraphics[width=4.5cm]{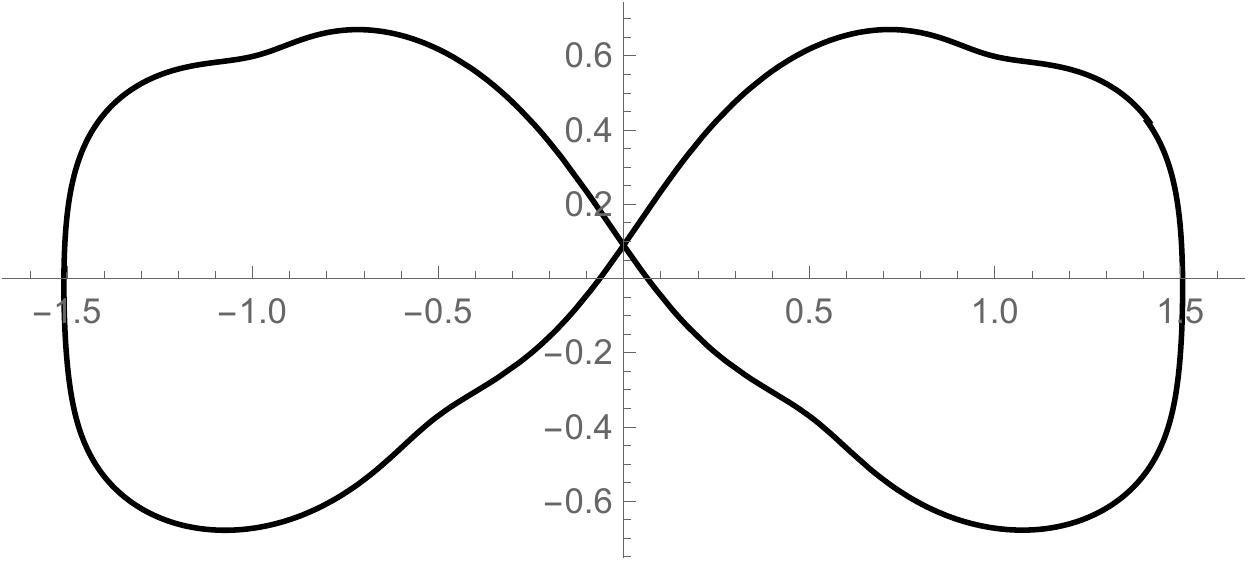}
	\\
	(a) \hspace*{4cm} (b) \hspace*{4cm} (c)
	\caption{
		Choreographic solutions for $T=20$;   
		(a) $C_x$ bifurcating from $\alpha_-$ at $T=14.595$, 
		(b) $C_2$ from $\alpha_+$ at $T=18.615$,
		(c) $C_y$ from $\alpha_+$ at $T=17.132$.
		$C_2$ is rotated by $\theta_2$ in (\ref{theta2}).
		Parameters; 
		(a) $(x, y, v, l)=(0.76038, 0.73802, 0.49633, 0.096581)$, 
		(b) $(x, u, v; \Delta I)=(1.4638, 0.11916, 0.30699; 0.48813)$,  
		(c) $(x, y, u, v, w, y')=(1.4167, 0.42115, -0.19613, 0.27560, -0.52564, 0.089593)$.
	}
	\label{Cx}
\end{figure}

At $T=14.595$ for $\alpha_-$,  
the variated orbit $Q$ is symmetric in the $x$ axis as shown in figure~\ref{Cx.fig}~(a).
We denote this orbit by $C_x$. %as in table \ref{LJ:tbl}.
Conditions for $q(t)$ to be $C_x$, derived in \ref{A.Cx}, are
that $q(t)$ takes an isosceles triangle configuration  
shown by filled circles in figure~\ref{Cx.fig}~(a), 
(\ref{q.tri}) and (\ref{p.tri}) with (\ref{D.tri.l}) 
at $t=0$ by parameters $(x,y,v,l)$, 
and opposite isosceles triangle configuration at $t=T/6$ 
shown by open circles in figure~\ref{Cx.fig}~(a), 
\begin{equation}\label{Cx.con}
	( q_6, \dot{q}_5, q_1-q_3, \dot{q}_2-\dot{q}_4 )=0. %, \; \dot{q}_5>0.
\end{equation}
%See figure~\ref{Cx.fig} (a).
%
For given period $T$, 
four conditions (\ref{Cx.con}) determine four parameters $(x, y, v, l)$. 
An index to distinguish solution $C_x$ from figure-eight choreography is $l \ne 0$. 
In figure~\ref{Cx}~(a), 
the orbit of the solution $C_x$ for $T=20$ 
bifurcated at $T=14.595$ from $\alpha_-$  is shown with parameters $(x,y,v,l)$.

At $T=18.615$ for $\alpha_+$,  
the variated orbit $Q$ is symmetric at origin as shown in figure~\ref{Cx.fig}~(b). 
We denote this orbit by $C_2$. % as in \cite{fukuda2}.
Conditions for $q(t)$ to be $C_2$, derived in \ref{A.C2}, are
that $q(t)$ takes an Euler configuration  
shown by filled circles in figure~\ref{Cx.fig}~(b), 
(\ref{q.Euler}) and (\ref{p.Euler}) at $t=0$ by parameters $(x,u,v)$, 
and another Euler configuration at $t=T/6$ 
shown by open circles in figure~\ref{Cx.fig}~(b), 
\begin{equation}\label{C2.con}
	( q_1, \, q_2, \dot{q}_1\dot{q}_6-\dot{q}_2\dot{q}_5)=0. %\; \dot{q}_3(t)>0.
\end{equation}
For given period $T$, 
three conditions (\ref{C2.con}) determine three parameters $(x, u, v)$. 
An index to distinguish solution $C_2$ from figure-eight choreography is $\Delta I \ne 0$.
In figure~\ref{Cx}~(b), the orbit of the solution $C_2$ for $T=20$ 
bifurcated at $T=18.615$ from $\alpha_+$ is shown with parameters $(x,y,v)$ and $\Delta I$.

At $T=17.132$ for $\alpha_+$,  
the variated orbit $Q$ is symmetric in the $y$ axis as shown in figure~\ref{Cx.fig}~(c).
We denote this orbit by $C_y$. % as in \cite{fukuda2}.
Conditions for $q(t)$ to be $C_y$, derived in \ref{A.Cy}, are 
that a body is on the $y$ axis at $t=0$ 
shown by filled circles in figure~\ref{Cx.fig}~(c), 
\begin{equation}\label{Cy.q}
	q(0)=(x,y, -x,-y-y', 0,y'),
\end{equation}
\begin{equation}\label{Cy.p}
	\dot{q}(0)=(u,v, -u-s,-v-w, s,w), 
\end{equation}
where
\begin{equation}\label{Cy.s}
	s = \frac{2(x v-y u)+x w-y' u}{2 y'+y}
\end{equation}
by parameters $(x,y, u,v, w,y')$,
and its inversion in the $y$ axis at $t=T/6$ 
shown by open circles in figure~\ref{Cx.fig}~(c), 
\begin{equation}\label{Cy.con}
	( q_1, q_3+x, q_4-y, \dot{q}_4+u, \dot{q}_5-v, q_2-y' )=0. %\; \dot{q}_3(t)>0.
\end{equation}
%See figure~\ref{Cx.fig} (c).
%
For given period $T$,   
the six conditions (\ref{Cy.con}) determine six parameters $(x,y, u,v, w,y')$. 
An index to distinguish solution $C_y$ from figure-eight choreography is $y' \ne 0$.
In figure~\ref{Cx}~(c), the orbit of the solution $C_y$ for $T=20$ 
bifurcating at $T=17.132$ from $\alpha_+$  is shown with $(x,y,u,v,w,y')$.

For these three choreographies $C=C_x, C_2, C_y$,
the variated orbit $Q$ in (\ref{q+psi:C}) is written as %for $C=C_x, C_2, C_y$ as
\begin{equation}\label{Qc}
	Q^C = q \pm h\psi^{(1)}.
\end{equation} 
The sign in front of $h$ effects inversion of the orbit in the $y$ axis for $C_x$, 
and in the $x$ axis for $C_2$ and $C_y$.  
These couples of congruent orbits bifurcate in the right side of the bifurcation point and $N_B=1$.

\section{Summary and discussions}
\label{sec:summary}
In this paper, 
we showed that the Morse index changes at a bifurcation point for periodic solution 
and inversely all points where the Morse index changes are bifurcation points
%for our example; 
for figure-eight choreography under homogeneous potential with $a \ge 0$ and 
for the $\alpha$ solutions under LJ-type potential.
Thus, for these choreographies, 
change of the Morse index, $\Delta N \ne 0$, 
is not only necessary but also sufficient condition for bifurcation point.
Further we observed that $\Delta N$ determines 
the number of incongruent bifurcating solutions $N_B$ as (\ref{N_B=DN}).

\begin{figure}
	\centering
	\includegraphics[width=6cm]{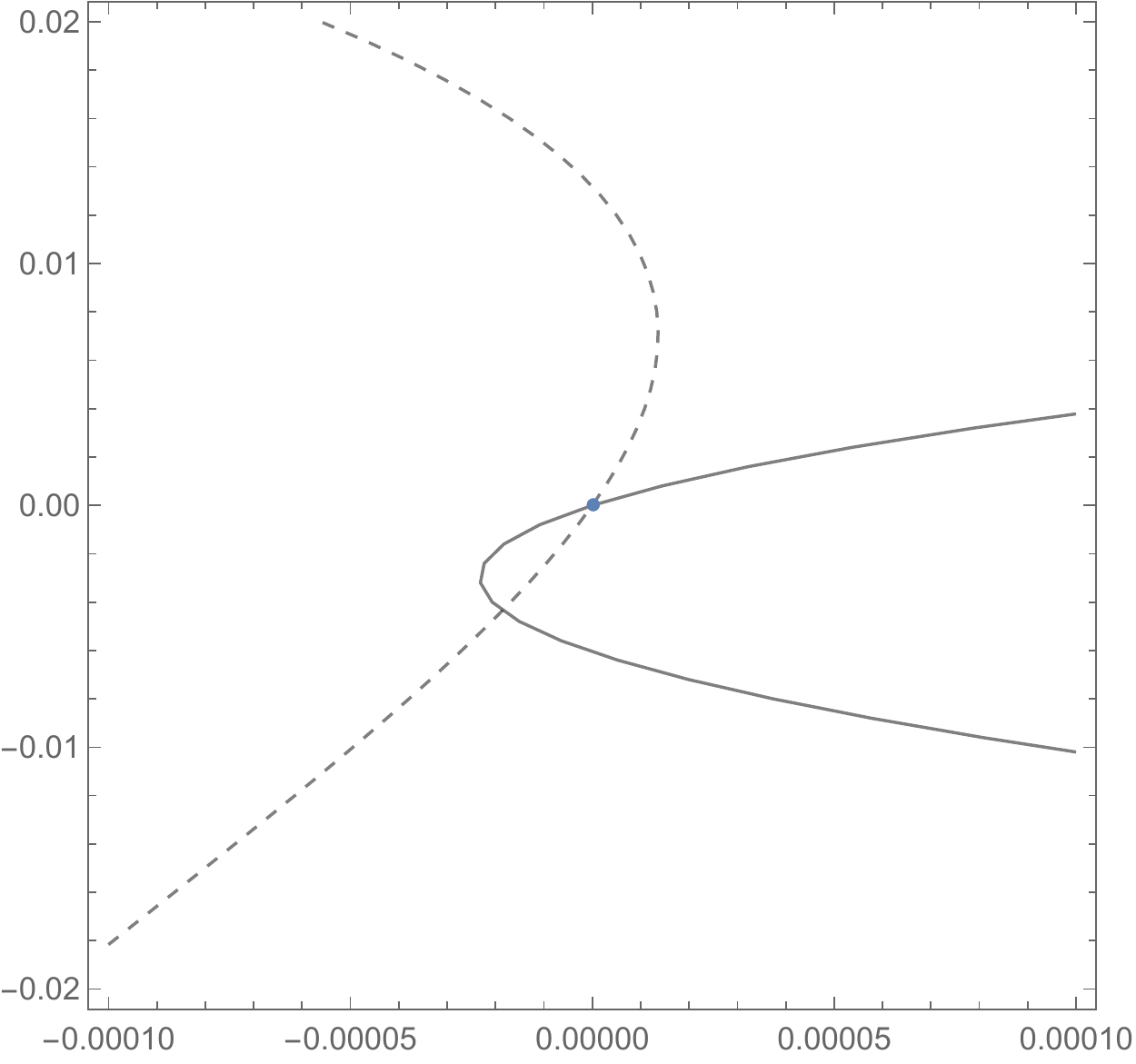} 
	\caption{
		Map to search $D_{x y}$ solution. 
		Full and dashed curves are $q_2(t)=0$ and 
		$\dot{q}_1(t) \dot{q}_6(t) - \dot{q}_2(t) \dot{q}_5(t)=0$, respectively,  
		at $t$ satisfying $q_4(t)=0$ for $q(t)$ starting from the initial conditions 
		(\ref{q.tri})--(\ref{H.theta}) with 
		$E=-0.69571$ for $a=1$ homogeneous potential.
		Horizontal and vertical axes are $x-x_0$ and $y-y_0$, respectively, 
		where $(x_0,y_0)$ is $(x,y)$ for figure-eight choreography with $T=1$. 
		Crossing points of two curves show 
		the parameters for the $D_{x y}$ solutions and for the figure-eight choreography.
	}
	\label{Hmap}
\end{figure}

The bifurcations are confirmed numerically by Newton's method.
If the number of parameters is three as in the $D_{x y}$, $D_2$ and $C_2$ cases,
the parameters of the solution are represented graphically like \cite{fukuda}.
%For $q(t)$ with the initial conditions (\ref{H.tri}), 
At $t$ satisfying one of the three conditions under some restriction $f=\mbox{const}$
where $f$ is a function of the three parameters,
%in three parameters $(x,y,v)$ is imposed then 
the rest conditions are given by two curves in the plane of two parameters. 
In figure~\ref{Hmap}, for solution $D_{x y}$ determined by three parameters $(x, y, v)$ 
in $a=1$ homogeneous system, 
at $t$ satisfying $q_4(t)=0$, the two conditions 
$q_2(t)=0$ and 
$\dot{q}_1(t) \dot{q}_6(t) - \dot{q}_2(t) \dot{q}_5(t)=0$
are shown in $(x,y)$ plane with total energy 
\begin{equation}\label{key}
	E = \frac{1}{2} \sum_{i=1}^6 \dot{q}_i^2 + U(q)
\end{equation}
as a restriction. 
The parameter $(x,y)$ for the $D_{x y}$ solution and the figure-eight choreography are 
observed as crossing points of two curves in figure~\ref{Hmap}.

\begin{figure}
	\centering
	\includegraphics[width=6cm]{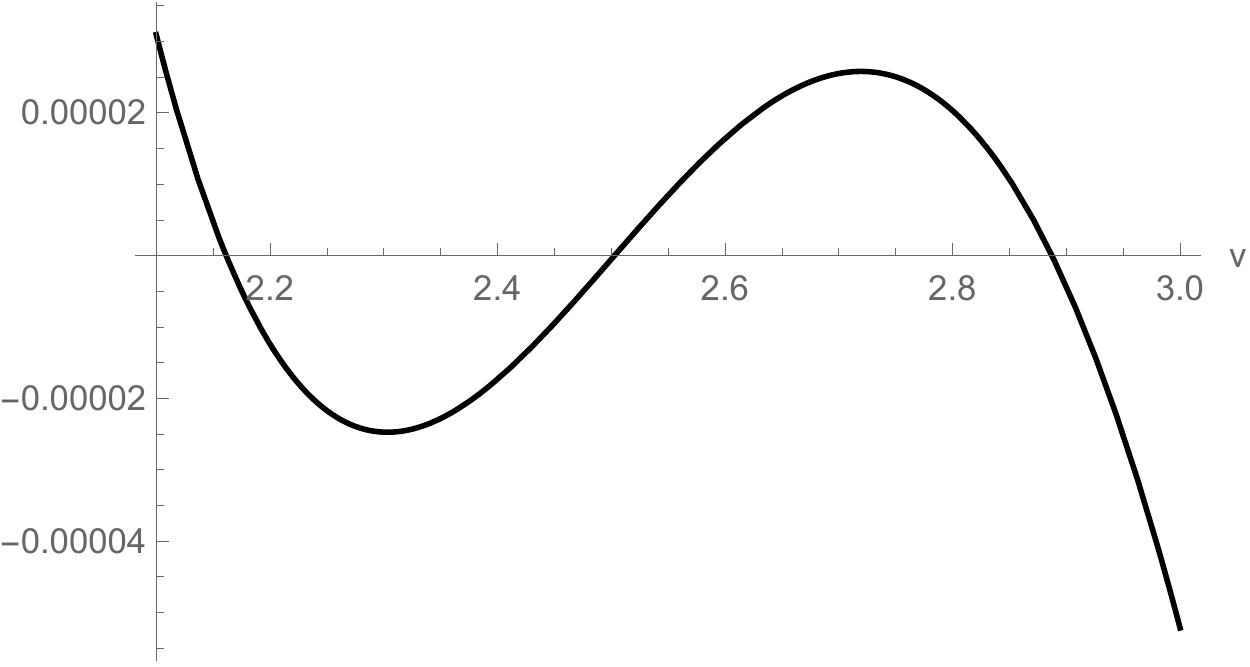} 
	\caption{
		$\dot{q}_3$ in (\ref{D.con}) as a function of $v$.
		zero about $v=2.5$ corresponds to figure-eight choreography and
		two other zeros bifurcating solution $D_x$.   
	}
	\label{q2dot}
\end{figure}

Newton's method sometimes does not converge unless initial parameters are good enough.
We introduced another graphically assisted method:
draw one condition as a function of one parameter by Newton's method in one lower dimension. 
In figure~\ref{q2dot}, 
for solution $D_x$ determined by four parameters $(x, y, v, l)$ in $a=1.3425$ homogeneous system, 
$\dot{q}_3$ in (\ref{D.con}) are shown as a function of $v$ 
which is obtained by $4-1$ dimensional Newton's method
for the rest of three parameters $(x,y,l)$
with the rest of three conditions $( q_4, q_1-q_5, \dot{q}_2-\dot{q}_6 )=0$. 
Zero about $v=2.5$ corresponds to figure-eight choreography and
two other zeros $D_x$ solutions in figure~\ref{q2dot}. 

It is useful to evaluate the Euler characteristics %$\chi$ 
\begin{equation}\label{chi:def}
	\chi = \sum_{q'} (-1)^{N(q')}
\end{equation} 
for the manifold of action functional in the domain of periodic functions,
where $N(q')$ is the Morse index at a periodic solution $q'$.
Since the $\chi$ will conserve, 
sometimes comparison of $\chi$ at the both sides of the bifurcation point 
helps to find bifurcating solution.

We assume the Euler characteristics conserves at the both sides of bifurcation point, and
(\ref{N_B=DN}) holds.  
Then we denote a number of congruent solutions 
which belong to the $i$th incongruent class by $n_i$, 
and their Morse index by $N_i$, $i=1,2,\ldots,N_B$. 

For a bifurcation point with $|\Delta N|=1$, 
bifurcation is one side since $N_B=1$. 
From the conservation of $\chi$ at the both sides of bifurcation point,  
\begin{equation}\label{key}
	(-1)^N = (-1)^{N \pm 1}+n_1 (-1)^{N_1}  
\end{equation}
we obtain $n_1=2$ and $(-1)^{N_1}=(-1)^N$. 

For a bifurcation point with $|\Delta N|=2$, 
bifurcation can be one side or both sides since $N_B=2$. 
For one side bifurcation,  
conservation of $\chi$,  
\begin{equation}\label{key}
	(-1)^N = (-1)^{N \pm 2}+n_1 (-1)^{N_1}+n_2 (-1)^{N_2}  
\end{equation}
leads $n_1=n_2$ and $(-1)^{N_1}=-(-1)^{N_2}$. 
For both sides bifurcation,   
\begin{equation}\label{key}
	(-1)^N+n_1 (-1)^{N_1} = (-1)^{N \pm 2}+n_2 (-1)^{N_2}  
\end{equation}
leads $n_1=n_2$ and $(-1)^{N_1}=(-1)^{N_2}$. 
In table \ref{Euler:tbl}, these restrictions on bifurcation are tabulated.
According to the restriction for one side bifurcation at $|\Delta N|=2$ 
we found $D_2$ solution after finding $D_x$.

\begin{table}%-------------------------------------------------------
	\centering
	\caption{
			Restrictions of bifurcation by conservation of the Euler characteristics and (\ref{N_B=DN}). 
		Symbols $n_i$ and $N_i$ are a number of congruent solutions 
		which belongs to the $i$th incongruent class, 
		and their Morse index, respectively, 
		$i=1,\ldots,N_B$, $N_B=|\Delta N|$.
	} 
	\begin{tabular}{c l l l}
		\hline
		$|\Delta N|$ & Bifurcation & Restriction for $n_i$ & Restriction for $N_i$ \\
		\hline
		$1$ & one side & $n_1=2$ & $(-1)^{N_1}=(-1)^N$ \\
		$2$ & one side & $n_1=n_2$ & $(-1)^{N_1}=-(-1)^{N_2}$ \\
		$2$ & both sides & $n_1=n_2$ & $(-1)^{N_1}=(-1)^{N_2}$ \\
		\hline
	\end{tabular}
	\label{Euler:tbl}
\end{table}%-------------------------------------------------------

The procedure to find bifurcation numerically by the Morse index is summarized as follows:
1) Find a point the Morse index changes, $\Delta N \ne 0$.
2) Investigate the corresponding variated orbit $Q$ %(\ref{q+psi}), 
which is an approximation of bifurcating solution. 
Indeed, symmetry of $Q$ is useful for numerical calculation.
3) If $|\Delta N| > 1$, variated orbit is chosen to make action critical.
4) Check conservation of the Euler characteristics at both sides of bifurcation point by table \ref{Euler:tbl}.

We leave the followings for future works: 
calculation of the Morse index for bifurcating solution; 
calculation of the Morse index for figure-eight choreography under homogeneous potential 
with negative $a$, that is, $r^{-a}$ for $a<0$; 
tracking solutions bifurcated as much as possible;
calculation of linear stability for figure-eight choreography and bifurcated solutions;
conditions for the point the Morse index changes to be bifurcation point;
and conditions for observation (\ref{N_B=DN})
on the number of bifurcating solutions to hold.

\ack
We thank Kazuyuki Yagasaki for his valuable comment on variated orbits at 
Symposium on Celestial Mechanics and $N$-body Dynamics (2017). 
This work was supported by JSPS Grant-in-Aid for Scientific Research 17K05146
(HF) and 17K05588 (HO).

\appendix
\section{Conditions for solutions}
We derive conditions for $q(t)$ to be 
$D_{x y}$, $D_x$, $D_2$, $C_x$, $C_2$ and $C_y$ solutions. 
We assume an inertia frame that 
total linear momentum $P$ is zero, $P=\sum_b \dot{\mathbf{r}}_b=0$, and 
center of mass $G$ is at origin, $G=\sum_b \mathbf{r}_b=0$.
%
%In the appendix 
The subscript and index for body 
is assumed to be in the range 1 and 3. % with translation by 3.

For $D_{x y}$, $D_2$, $C_2$ and $C_y$, total angular momentum 
$l=\sum_b \mathbf{r}_b \times \dot{\mathbf{r}}_b$ is zero
since sum of signed area of three orbits is zero from the symmetry of orbits.

A configuration 
that a body $b$ is in the $x$ axis and the other two bodies $b \pm 1$ have 
the same $x$ coordinate 
is called isosceles triangle configuration. 
For $q(t)$ symmetric in the $x$ axis
conditions for isosceles triangle configuration is written as
\begin{equation}\label{A.tri}
	( q_{2 b}, \dot{q}_{2 b-1}, q_{2 b+1}-q_{2 b-3}, \dot{q}_{2 b+2}-\dot{q}_{2 b-2} )=0
\end{equation}
by $P=0$. 
For $b=2$ 
with $x=q_1$, $y=q_2$, $v = \sqrt{\dot{q}_1^2+\dot{q}_2^2}$ 
and total angular momentum $l$,  (\ref{A.tri}) is written
as (\ref{q.tri}), (\ref{p.tri}) and (\ref{D.tri.l}). 
The motion beginning from (\ref{A.tri})
is time reversal motion  from (\ref{A.tri}) 
with inversion in the $x$ axis and exchange of bodies $b \pm 1$.

A configuration 
that a body $b$ is at origin is called as Euler configuration.
For $q(t)$ symmetric at origin
a condition for Euler configuration is written as
\begin{equation}\label{A.Euler}
	( q_{2 b-1}, q_{2 b}, \dot{q}_{2 b+1} \dot{q}_{2 b-2} - \dot{q}_{2 b+2} \dot{q}_{2 b-3} )=0
\end{equation}
by $G=0$.
For $b=3$ with $x=q_1$, $u=\dot{q}_1$, $v=\dot{q}_2$ 
(\ref{A.Euler}) is written as (\ref{q.Euler}) and (\ref{p.Euler}). 
The motion beginning from (\ref{A.Euler})
is time reversal motion from (\ref{A.Euler}) 
with $\pi$ rotation and exchange of bodies $b \pm 1$.

\subsection{$D_{x y}$ solution}
\label{A.H}

The $D_{x y}$ solution takes initial conditions (\ref{q.tri}) and (\ref{p.tri}) with (\ref{H.theta}) 
and takes Euler configuration when body 2 reaches at origin at $t=t'$.
Thus (\ref{A.Euler}) with $b=2$, (\ref{H.Euler}) holds at $t=t'$.
%Thus $\pi$ rotation and exchange of bodies 1 and 3 is equivalent to time reversal at $t=t'$.
%Thus motion from $t=t'$ to $t=2 t'$ is time reversal and $\pi$ rotation and exchange of bodies 1 and 3.
Thus at $t=2 t'$, three bodies take initial conditions 
with $\pi$ rotation and exchange of bodies 1 and 3.
Then at $t=4 t'$,  
%Then motion from $t=t'$ to $t=2 t'$ is that from $t=0$ to $t'$ with 
%$\pi$ rotation and exchange of bodies 1 and 3, thus at $t=4 t'$, 
three bodies take initial conditions again 
and the motion is periodic with period $T=4 t'$. %then $t'=T/4$. 
Since the orbits were constructed 
by those from $t=0$ to $t=T/2$ and their $\pi$ rotation, 
the orbits are symmetric in the $x$ and the $y$ axes.

\subsection{$D_x$ solution}
\label{A.Dx}

The $D_x$ solution takes initial conditions (\ref{q.tri}) and (\ref{p.tri}) with (\ref{H.theta}) 
and takes isosceles triangle configuration when body 2 reaches in the $x$ axis again at $t=t'$.
Then (\ref{A.tri}) with $b=2$, (\ref{D.tri.l}) holds at $t=t'$.
Then 
at $t=2 t'$, three bodies take initial conditions again
and the motion is periodic with period $T=2 t'$. %then $t'=T/2$. 
Since the orbits were constructed 
by those from $t=0$ to $t=T/2$ and their inversion in the $x$ axis, 
the orbits are symmetric in the $x$ axis.

\subsection{$C_x$ solution}
\label{A.Cx}

The $C_x$ solution takes initial conditions (\ref{q.tri}) and (\ref{p.tri}) with (\ref{H.theta}) 
and takes isosceles triangle configuration when body 3 reaches in the $x$ axis again at $t=t'$.
Then (\ref{A.tri}) with $b=3$, (\ref{Cx.con}) holds at $t=t'$.
Then at $t=2 t'$, three bodies take initial conditions again but with cyclic permutation of bodies, 
and the motion is choreographic with period $T=3 \times 2 t'=6 t'$. % then $t'=T/6$. 
Since the orbits were constructed 
by those from $t=0$ to $t=T/6$ and their inversion in the $x$ axis,
the orbits are symmetric in the $x$ axis.
 
\subsection{$D_2$ solution}
\label{A.D2}

The $D_2$ solution takes initial conditions (\ref{q.Euler}) and (\ref{p.Euler}), 
and takes Euler configuration when body 2 reaches at origin at $t=t'$.
Then (\ref{A.Euler}) with $b=2$, (\ref{D2.con}) holds at $t=t'$.
Thus 
at $t=2 t'$, three bodies take initial conditions again
and the motion is periodic with period $T=2 t'$. % then $t'=T/2$. 
Since the orbits were constructed 
by those from $t=0$ to $t=T/2$ and their $\pi$ rotation, 
the orbits are symmetric at origin.

\subsection{$C_2$ solution}
\label{A.C2}

The $C_2$ solution takes initial conditions (\ref{q.Euler}) and (\ref{p.Euler}), 
and Euler configuration when body 1 reaches in the $x$ axis again at $t=t'$.
Then (\ref{A.tri}) with $b=1$, (\ref{C2.con}) holds at $t=t'$.
Thus at $t=2 t'$, three bodies take initial conditions again with cyclic permutation of bodies, 
and the motion is choreographic with period $T=3 \times 2 t'=6 t'$. % then $t'=T/6$. 
Since the orbits were constructed 
by those from $t=0$ to $t=T/6$ and their $\pi$ rotation,
the orbits are symmetric at origin.

\subsection{$C_y$ solution}
\label{A.Cy}

The $C_y$ solution takes initial conditions (\ref{Cy.q}) and (\ref{Cy.p}) by $G=P=0$ 
with (\ref{Cy.s}) by $l=0$ where body 3 is on the $y$ axis, 
which are represented by six parameters.
At $t=t'$ when body 1 reaches in the $y$ axis, 
the relation between the positions and velocities at $t=t'$ % and $0$ 
have to be inversion in the $y$ axis with exchange of bodies 2 and 3.   
Thus we have twelve relations but six conservation quantities, $G$, $P$, $l$ 
and total energy, reduce it to six as (\ref{Cy.con}). 
Thus at $t=2 t'$ if (\ref{Cy.con}) is satisfied, 
three bodies take initial conditions again with cyclic permutation of bodies, 
and the motion is choreographic with period $T=3 \times 2 t'=6 t'$. % then $t'=T/6$. 
Since the orbits were constructed 
by those from $t=0$ to $t=T/6$ and their inversion in the $y$ axis, 
the orbits are symmetric in the $y$ axis.

%\References

\end{document}